\documentclass[12pt,preprint]{aastex}

\usepackage{natbib}
\usepackage[utf8]{inputenc}
\usepackage{graphicx}
\usepackage{amsmath}
\usepackage{lscape}

\begin{document}

\title{
A Subarcsecond ALMA Molecular Line Imaging Survey of the Circumbinary, Protoplanetary Disk Orbiting V4046 Sgr} 

\author{Joel H. Kastner\altaffilmark{1,2}, C. Qi\altaffilmark{3}, D. A. Dickson-Vandervelde\altaffilmark{2},
  P. Hily-Blant\altaffilmark{4,5}, T. Forveille\altaffilmark{4}, S. Andrews\altaffilmark{3}, U. Gorti\altaffilmark{6}, K. \"{O}berg\altaffilmark{3}, D. Wilner\altaffilmark{3}}

\altaffiltext{1}{Chester F. Carlson Center for Imaging Science, Rochester Institute of
Technology, 54 Lomb Memorial Drive, Rochester NY 14623 USA
(jhk@cis.rit.edu)}
\altaffiltext{2}{School of Physics \& Astronomy and 
Laboratory for Multiwavelength Astrophysics, Rochester Institute of
Technology, 54 Lomb Memorial Drive, Rochester NY 14623 USA}
\altaffiltext{3}{Harvard-Smithsonian Center for Astrophysics, 60 Garden Street, Cambridge, MA 02138, USA}
\altaffiltext{4}{Universit\'{e} Grenoble Alpes, Institut de Plan\'{e}tologie et d'Astrophysique de Grenoble (IPAG), F-38000, Grenoble, France; CNRS, IPAG, F-38000, Grenoble, France}
\altaffiltext{5}{Institut Universitaire de France, F-38000, Grenoble, France}
\altaffiltext{6}{SETI Institutute, NASA Ames Research Center, Mountain View, CA, USA}

\begin{abstract}
We present a suite of ALMA interferometric molecular line and continuum images that elucidate, on linear size scales of $\sim$30--40 AU, the chemical structure of the nearby, evolved, protoplanetary disk orbiting the close binary system V4046 Sgr. The observations were undertaken in the 1.1--1.4 mm wavelength range (ALMA Bands 6 and 7) with antenna configurations involving maximum baselines of several hundred meters, yielding subarcsecond-resolution images in more than a dozen molecular species and isotopologues. Isotopologues of CO and HCN display centrally peaked morphologies of integrated emission line intensity, whereas the line emission from complex nitrile group molecules (HC$_3$N, CH$_3$CN), deuterated molecules (DCN, DCO$^+$), hydrocarbons (as traced by C$_2$H), and potential CO ice line tracers (N$_2$H$^+$, and H$_2$CO) appears as a sequence of sharp and diffuse rings. The dimensions and morphologies of HC$_3$N and CH$_3$CN emission are suggestive of photodesorption of organic ices from the surfaces of dust grains, while the sequence of increasing radius of peak intensity represented by DCN (smallest), DCO$^+$, N$_2$H$^+$, and H$_2$CO (largest) is qualitatively consistent with the expected decline of midplane gas temperature with increasing disk radius. Empirical modeling indicates that the sharp-edged C$_2$H emission ring lies at relatively deep disk layers, leaving open the question of the origin of C$_2$H abundance enhancements in evolved disks. This study of the ``molecular anatomy'' of V4046 Sgr should serve as motivation for additional subarcsecond ALMA molecular line imaging surveys of nearby, evolved protoplanetary disks aimed at addressing major uncertainties in protoplanetary disk physical and chemical structure and molecular production pathways. 
\end{abstract}

\section{Introduction}

Contemporary models describing viscously heated, irradiated protoplanetary disks orbiting solar-mass pre-main sequence (T Tauri) stars typically invoke a combination of gas-phase, gas-grain, and grain surface processes, some of which are driven by the intense dissociating and ionizing radiation from the central stars \citep[e.g.,][]{Cleeves2013,Walsh2015}. The physical and chemical disk structures that emerge from this complex admixture at late disk evolutionary stages give rise to the formation of
molecular disk ``ice lines'' and ``dead zones'' and, hence, ultimately determine
the semimajor orbits, masses, and compositions of any resulting planets \citep{ObergBergin2016,Cridland2017}. 
At late disk evolutionary stages these processes are accompanied by planet-disk (dynamical) interactions and radiation-driven disk photoevaporation, generating steep gradients in density, molecular abundance, and dust grain size that manifest themselves in the form of disk holes, rings, and gaps  \citep[e.g.,][]{Gorti2015,DongFung2017}. 

The handful of examples of nearby ($D\stackrel{<}{\sim}100$ pc), evolved (age $\sim$5--20 Myr) pre-main sequence stars of roughly solar mass that are orbited by and actively accreting from primordial circumstellar
disks offer unparalleled opportunities to investigate these and other late-stage planet formation processes \citep{Kastner2016IAUScont}. Two of these star-disk systems, TW Hya and V4046 Sgr  --- which lie at $D = 60.09\pm0.14$ pc and $72.47\pm0.34$ pc, respectively \citep{Gaia2018} --- have been among the most popular subjects for early mm-wave interferometric imaging studies of disks carried out with the Atacama Large Millimeter Array (ALMA). These initial ALMA studies are yielding new insight into the late physical and chemical evolution of protoplanetary disks \citep[e.g.,][]{Qi2013,Andrews2016, Bergin2016, Nomura2016, Huang2017, Guzman2017, HilyBlant2017,Huang2018}. 

Although TW Hya is thus far (and by far) the more heavily scrutinized of these two nearby, evolved disks, V4046 Sgr is in certain respects even more interesting. A member of the $\sim$23 Myr-old $\beta$ Pic Moving Group \citep{Mamajek2016}, V4046 Sgr is even older than TW Hya \citep[age $\sim$8 Myr;][]{Ducourant2014}; furthermore, V4046 Sgr is a close ($P\sim 2.4$ day) binary system consisting of nearly equal-mass, $\sim$0.9 $M_\odot$ components \citep[][and refs.\ therein]{Donati2011,Rosenfeld2012}. Despite its advanced system age, V4046 Sgr is orbited by a chemically rich circumbinary disk \citep{Kastner2008,Kastner2014,Rapson2015b} whose radial extent \citep[$\sim$350 AU;][]{Rodriguez2010} and estimated gas mass \citep[$\sim$0.09 $M_\odot$;][]{Rosenfeld2013} are even larger than those of the TW Hya molecular disk \citep[$\sim$200 AU and $\sim$0.05 $M_\odot$, respectively;][]{Andrews2012,Bergin2013}.  Thus, the V4046 Sgr system presents astronomers with a prime target to exploit so as to improve our understanding of the process of circumbinary planet formation around near-solar-mass stars in tight orbits.

Our single-dish molecular line surveys have established that the TW Hya and V4046 Sgr disks display remarkably similar molecular spectra, with particularly strong emission from HCO$^+$, HCN, CN, and C$_2$H in addition to CO \citep[][]{Kastner1997,Kastner2008,Kastner2014}. However, the two disks have sharply contrasting submm continuum emission morphologies: the submm surface brightness of TW Hya is centrally peaked, with a sharp outer edge at $\sim$60 AU and a set of superimposed gaps \citep[][]{Andrews2016}, whereas V4046 Sgr appears as a compact ring that peaks at a radius of $\sim$30 AU \citep{Rosenfeld2013}. The inner radius of this ring-like submm continuum emission structure is far too large to be the result of dynamical interactions between the disk and central binary, and is instead the hallmark of a ``transition disk'' --- i.e., a disk that has developed an inner cavity that is devoid of large dust grains as a consequence of its relatively advanced evolutionary state \citep[e.g.,][and references therein]{Andrews2011}. 

As in the case of TW Hya \citep{Rapson2015c}, scattered-light near-IR observations of V4046 Sgr with Gemini Planet Imager (GPI) revealed a dust ring system within the giant planet formation region of the 
disk \citep{Rapson2015a}. The outer (radius 30--45
AU) faint scattering halo imaged in the near-IR overlaps the ring detected in submm continuum
imaging of the disk, while the inner (radius
12--18 AU) bright ring seen in scattered light lies fully interior to the submm emission
``hole,'' demonstrating that the apparent (submm) cavity within $\sim$30 AU is actually rich in small
grains that have ``filtered'' through to smaller radii. When considered in the context of theoretical predictions of the effects of planets on disk structure \citep[e.g.,][]{Rice2003,PaardekooperMellema2006,Zhu2012}, the combination of a disk ``gap'' near 20 AU, an inner disk hole, and evidence for grain size segregation is indicative of recent or ongoing giant planet building activity \citep{Rapson2015a}. 

Initial ALMA disk studies have included V4046 Sgr among small samples of objects that were the subjects of molecular line imaging surveys focused on disk deuterium chemistry \citep{Huang2017}, nitrogen isotopic ratios \citep{Guzman2017}, and molecular ice lines (C. Qi et al.\ 2017, in preparation). In addition, we recently carried out an ALMA line imaging study of V4046 Sgr that was aimed at investigating the origin of the large CN and C$_2$H abundances that are evidently characteristic of evolved molecular disks \citep[see, e.g.,][]{Kastner2014,Punzi2015,Kastner2015,Bergin2016}. All of these ALMA line surveys of V4046 Sgr have been undertaken in the 1.1--1.4 mm wavelength range (ALMA Bands 6 and 7) with antenna configurations involving maximum baselines of several hundred meters, yielding subarcsecond-resolution images in more than a dozen molecular species and isotopologues. Here, we bring together the results of these observations of the disk orbiting V4046 Sgr. Collectively, these ALMA images serve to elucidate, on linear size scales of $\sim$30--40 AU, the chemical structure of an evolved, circumbinary, protoplanetary disk. 

\section{Observations}

The library of ALMA images of V4046 Sgr presented here was compiled from data obtained during the course of ALMA programs
2013.1.00226.S (Cycle 2; PI: K. \"{O}berg), 2015.1.00671.S (Cycle 3; PI: J. Kastner), and 2015.1.00678.S (Cycle 3; PI: C. Qi).
Table \ref{tbl:LibraryImages} lists the species and transitions imaged during the course of these programs, as well as the spatial and spectral resolution and sensitivity achieved for observations of each molecular isotopologue. The details of data acquisition for program 2013.1.00226.S have previously been described in papers by \citet{Huang2017}, who presented analysis of lines of deuterated and C isotopologues of HCO$^+$ and HCN, and \citet{Guzman2017}, who presented analysis of N  isotopologues of HCN. Additional molecular line images from program 2013.1.00226.S in lines of HC$_3$N and CH$_3$CN are presented in Bergner et al.\ (2018), while line images in isotopologues of CO are presented here for the first time. 

Data for program 2015.1.00671.S were obtained in an ALMA band 6 spectral setup optimized for the HCN $J=3 \rightarrow 2$ and C$_2$H $N=3 \rightarrow 2$ transitions. These observations were carried out during a 613 s integration with 41 antennas, in a configuration with baselines in the range 15--704 m, on 20 June 2016\footnote{Data obtained for program 2015.1.00671.S during a shorter ($\sim$360 s) integration on 31 July 2016 in a second spectral setup optimized for the CO $J=2 \rightarrow 1$ and CN $N=2 \rightarrow 1$ transitions (in the 226--231 GHz frequency range) could not be properly calibrated and are hence unusable.}. The central frequency, bandwidth, and channel widths were 262.9 GHz, 2474 MHz, and 231.9 kHz, respectively, for the C$_2$H $N=3 \rightarrow 2$ transition and 265.9 GHz, 618.6 MHz, and 116 kHz for the HCN $J=3 \rightarrow 2$ transition. The bandwidth and channel widths for the C$_2$H spectral setup were broader so as to include all hyperfine transitions in the bandpass. The bandpass, phase, and flux spectral calibrators were J1924$-$2914, J1816$-$3052, and J1733$-$1304, respectively. 

Data for program 2015.1.00678.S were collected in a 1200 s integration on 2016 April 30. The configuration consisted of 41 antennae with baselines in the range 20--640 m. The setup consisted of seven Band 7 spectral windows: four with spectral resolution of 244 kHz and bandwidths of 234 MHz centered on the $J=4 \rightarrow 3$ transitions of DCO$^+$ and DCN and the CH$_3$OH $6_{0,6}-5_{0,5}$ and H$_2$CO $4_{0,4}-3_{0,3}$ transitions; two with spectral resolution of 122 kHz and bandwidths of 117 MHz centered on the N$_2$H$^+$ $J=3 \rightarrow 2$ and CH$_3$OH 9$_{1,9} \rightarrow 8_{0,8}$ transitions; and one with resolution of 977 kHz and bandwidth of 938 MHz, for continuum. Here, we present the DCN, H$_2$CO, and N$_2$H$^+$ images; analysis of the other line image data will be presented in a forthcoming paper (Qi et al., in prep.).

Data for all three programs were initially calibrated by ALMA/NAASC staff. We then performed additional calibration, processing, and imaging using standard tools available in CASA 4.7.0. 
This basic self-calibration and imaging strategy is described in detail in \citet{Huang2017}, who first presented the DCO$^+$ and H$^{13}$CO data from program 2013.1.00226.S reanalyzed here.
To summarize, the continuum visibilities were extracted by averaging the line-free channels in each spectral window. We  then performed phase and amplitude self-calibration on the continuum visibilities. The same self-calibration solutions were applied to the channels containing line emission, after subtracting the average continuum obtained from line-free channels. The continuum and line images were reconstructed from the visibilities using Briggs \verb+robust+ parameters ranging from 0.0 to 2.0, with values chosen to obtain beamwidths of 0.5--0.6$''$ (see Table \ref{tbl:LibraryImages}, footnote {\it a}). 
The line data were cleaned per channel using a uniform elliptical clean mask encompassing the emission over all channels. We then used the resulting image cubes to generate the integrated intensity (moment 0) and intensity-weighted mean radial velocity (moment 1) images discussed in \S 3. 

The resulting clean beam sizes and integrated fluxes (within apertures encompassing the source flux above $\sim$3 $\sigma$ levels) are listed in Table \ref{tbl:LibraryImages} and Table \ref{tbl:Continuum} for the line and continuum imaging, respectively. The formal uncertainties on the line fluxes listed in Table \ref{tbl:LibraryImages} were estimated by propogating the rms per-channel errors, accounting for the radial extent of the emission in the moment 0 image as well as the linewidths \citep[see, e.g.,][]{Kastner2010}; the minimum uncertainties are $\sim$10\%, after accounting for typical systematic ALMA calibration errors. Our integrated continuum and $^{12}$CO and $^{13}$CO line flux measurements are consistent with those obtained in previous SMA imaging of V4046 Sgr within the respective uncertainties \citep{Rodriguez2010,Oberg2011}, with the possible exception of the 264 GHz continuum measurement \citep[which is $\sim$20\% lower than previously measured by][]{Oberg2011}.

\section{Results}

\subsection{Images}

\begin{figure}[htbp]
\begin{center}
\includegraphics[width=6.5in]{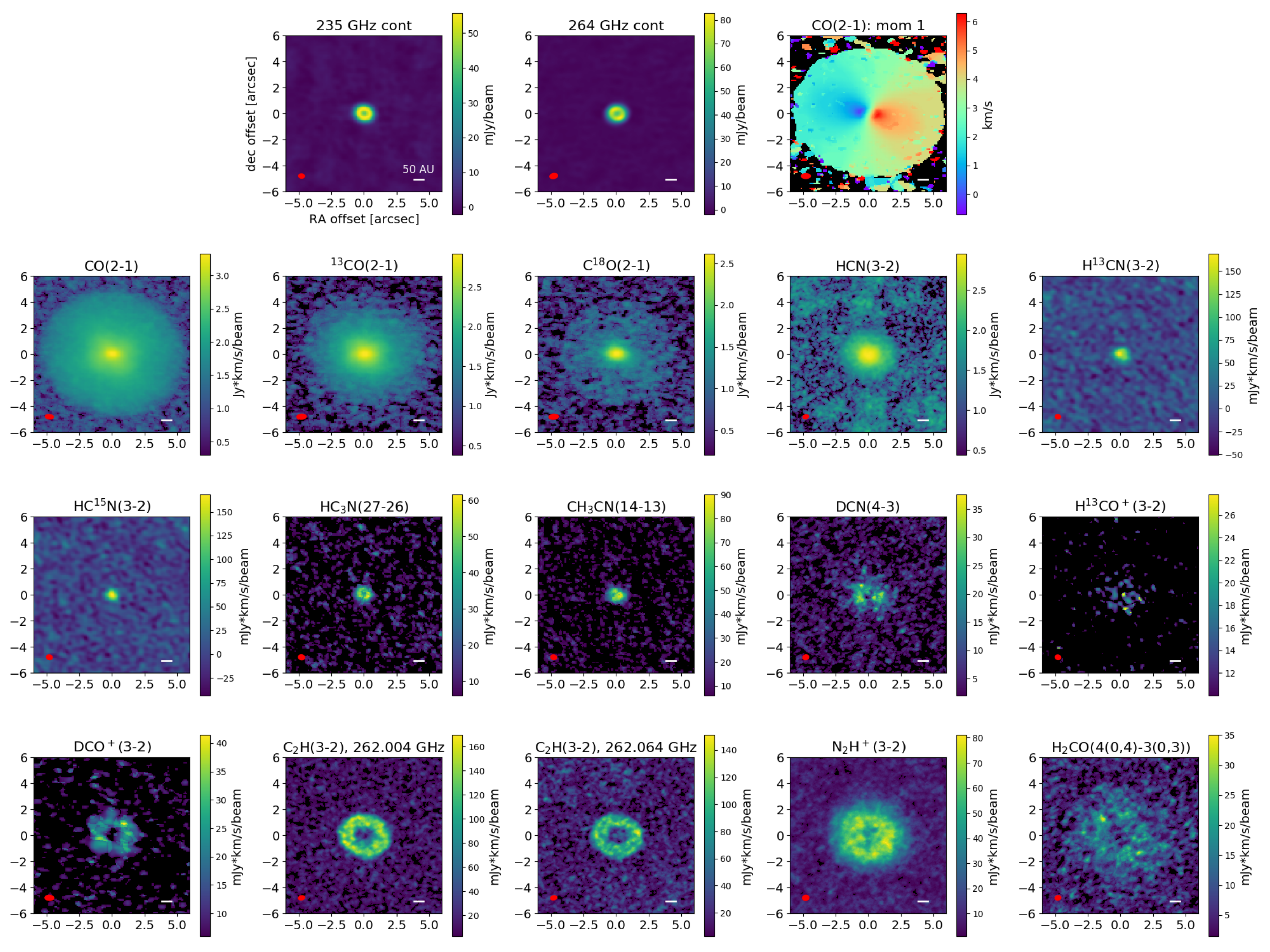}
\caption{Subarcsecond resolution ALMA images of the V4046 Sgr disk in continuum emission at 235 GHz and 260 GHz and in the molecular transitions listed in Table~\ref{tbl:LibraryImages}, displayed in $12''\times12''$  fields. Images are displayed with linear scaling, except for $^{12}$CO, $^{13}$CO, C$^{18}$O, and HCN, which use log scaling. The $^{12}$CO moment 1 image is at top right.
The red ellipse in each frame indicates the beam size and orientation.
}
\label{fig:AllImages}
\end{center}
\end{figure}

\begin{figure}[htbp]
\begin{center}
\includegraphics[width=6.5in]{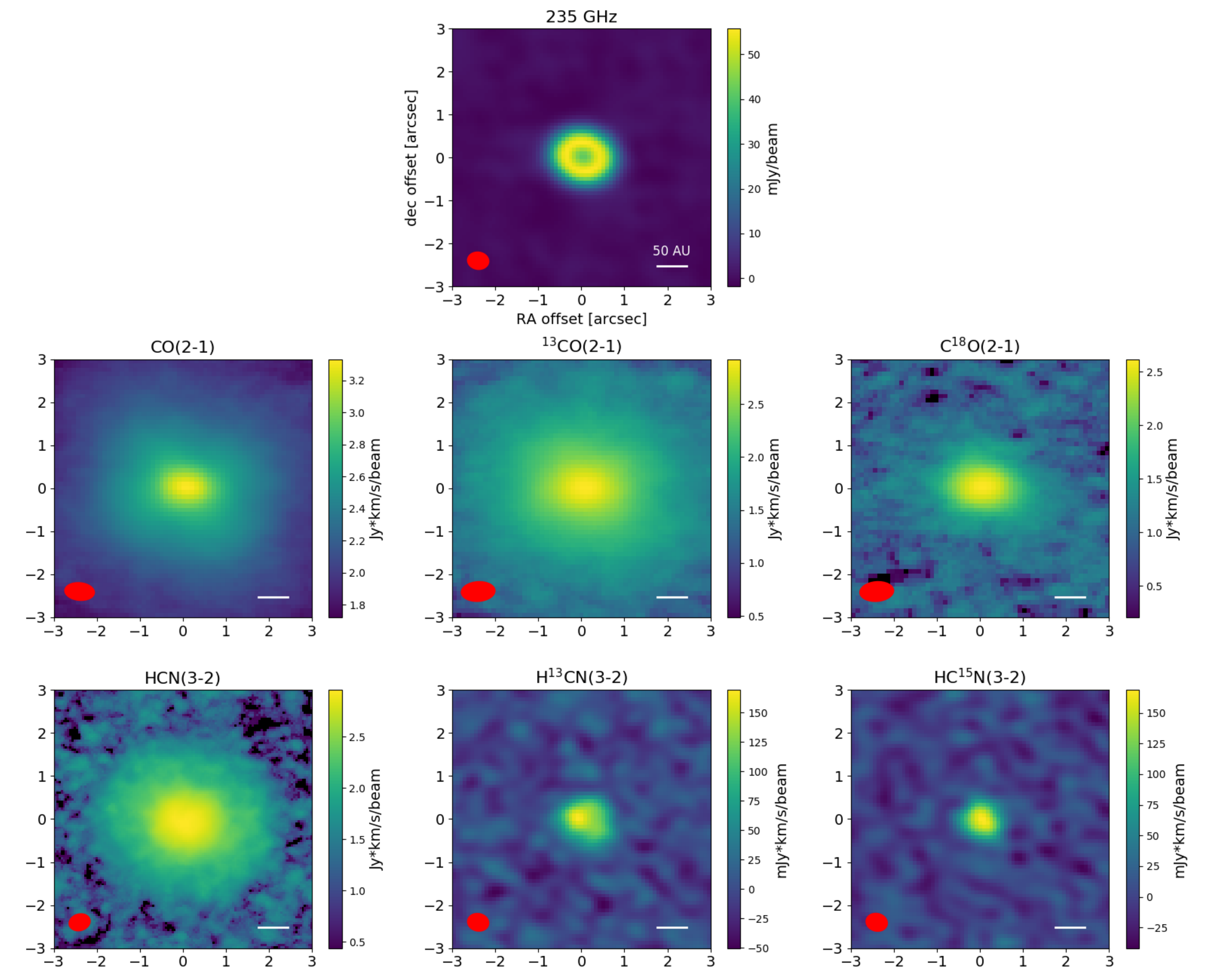}
\caption{As in Fig.~\ref{fig:AllImages} for emission-line images of CO, HCN, and isotopologues, all of which display centrally peaked emission morphologies; images are displayed in $6''\times6''$  fields. 
}
\label{fig:CPImages6by6}
\end{center}
\end{figure}

\begin{figure}[htbp]
\begin{center}
\includegraphics[width=6.5in]{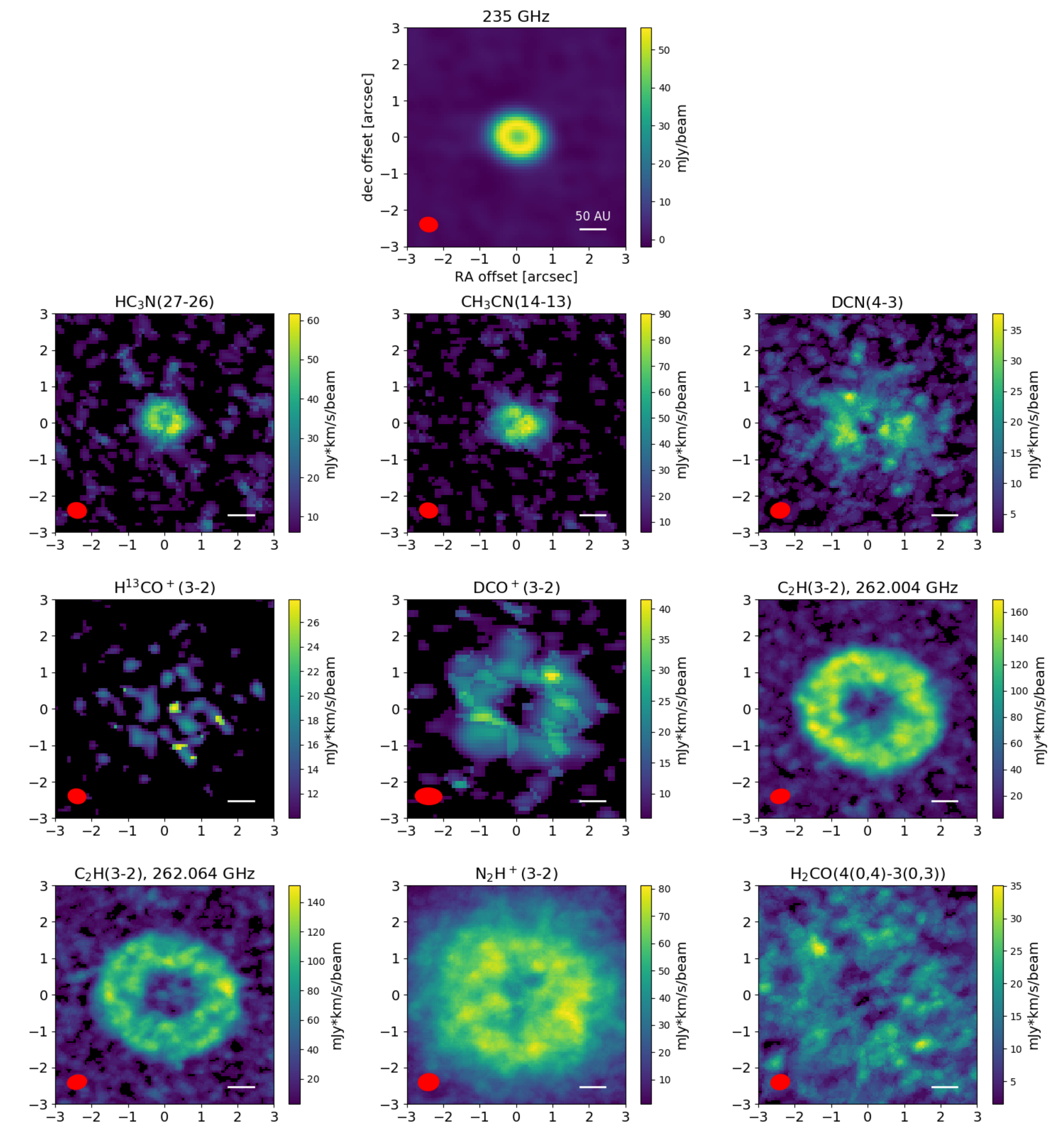}
\caption{As in Fig.~\ref{fig:CPImages6by6}, for emission-line images of species that display ring-like emission morphologies. 
}
\label{fig:RingImages6by6}
\end{center}
\end{figure}

The resulting ALMA 235 GHz and 264 GHz continuum images 
and moment 0 line images for the molecular species and transitions listed in Table \ref{tbl:LibraryImages}, as well as a moment 1 $^{12}$CO(2--1) image, are presented in Fig.~\ref{fig:AllImages} in $12''\times12''$ fields of view and in  Figs.~\ref{fig:CPImages6by6} and \ref{fig:RingImages6by6} in $6''\times6''$ fields of view. The moment 0 images displayed in these figures have been integrated over channel ranges that span the detectable emission; for all images apart from H$_2$CO and N$_2$H$^+$, the component channel maps were clipped at the 1$\sigma$ level, so as to isolate the signal. 

The continuum, $^{12}$CO, and $^{13}$CO images reproduce, at higher spatial resolution and signal-to-noise, the basic features seen in the best previous 1.3 mm continuum and CO imaging of V4046 Sgr, which were obtained with the SMA and presented in \citet{Rosenfeld2013}. Specifically, the mm-wave continuum emission appears as a partially filled ring, peaking at $\sim$0.3$''$ ($\sim$22 AU) from the star, with a sharp intensity cutoff at its outer radius of $\sim$0.8$''$ ($\sim$60 AU), whereas the CO is strongly centrally peaked and the $^{12}$CO  emission outer radius ($\sim$4$''$, i.e., $\sim$300 AU) is far larger than that of the continuum. As was the case  for the previous SMA imaging \citep{Rodriguez2010,Rosenfeld2013}, the large-scale Keplerian rotation of the disk is readily apparent in the moment 1 image of $^{12}$CO(2--1). 

\begin{figure}[htbp]
\begin{center}
\includegraphics[width=4.5in]{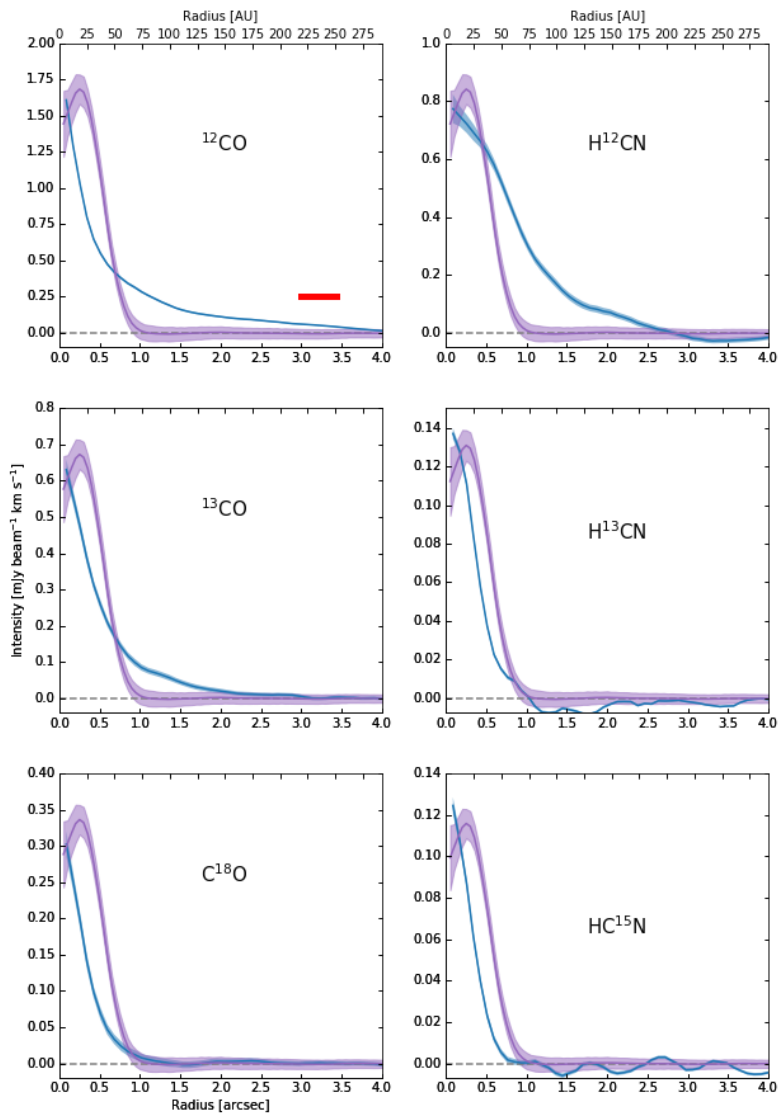}
\caption{Radial profiles extracted from integrated intensity emission-line images of  isotopologues of CO and HCN, all of which display centrally peaked morphologies. The appropriately normalized radial profile of 264 GHz continuum emission is overlaid (as a purple curve) in each panel, for reference. The red bar in the upper left panel indicates a beamwidth of $0.6''$, which is representative of the line survey data (i.e., beam major axes ranging from $\sim$0.52$''$ to $\sim$0.8$''$; Table~1).
}
\label{fig:CPradialProfiles}
\end{center}
\end{figure}

\begin{figure}[htbp]
\begin{center}
\includegraphics[width=4.0in]{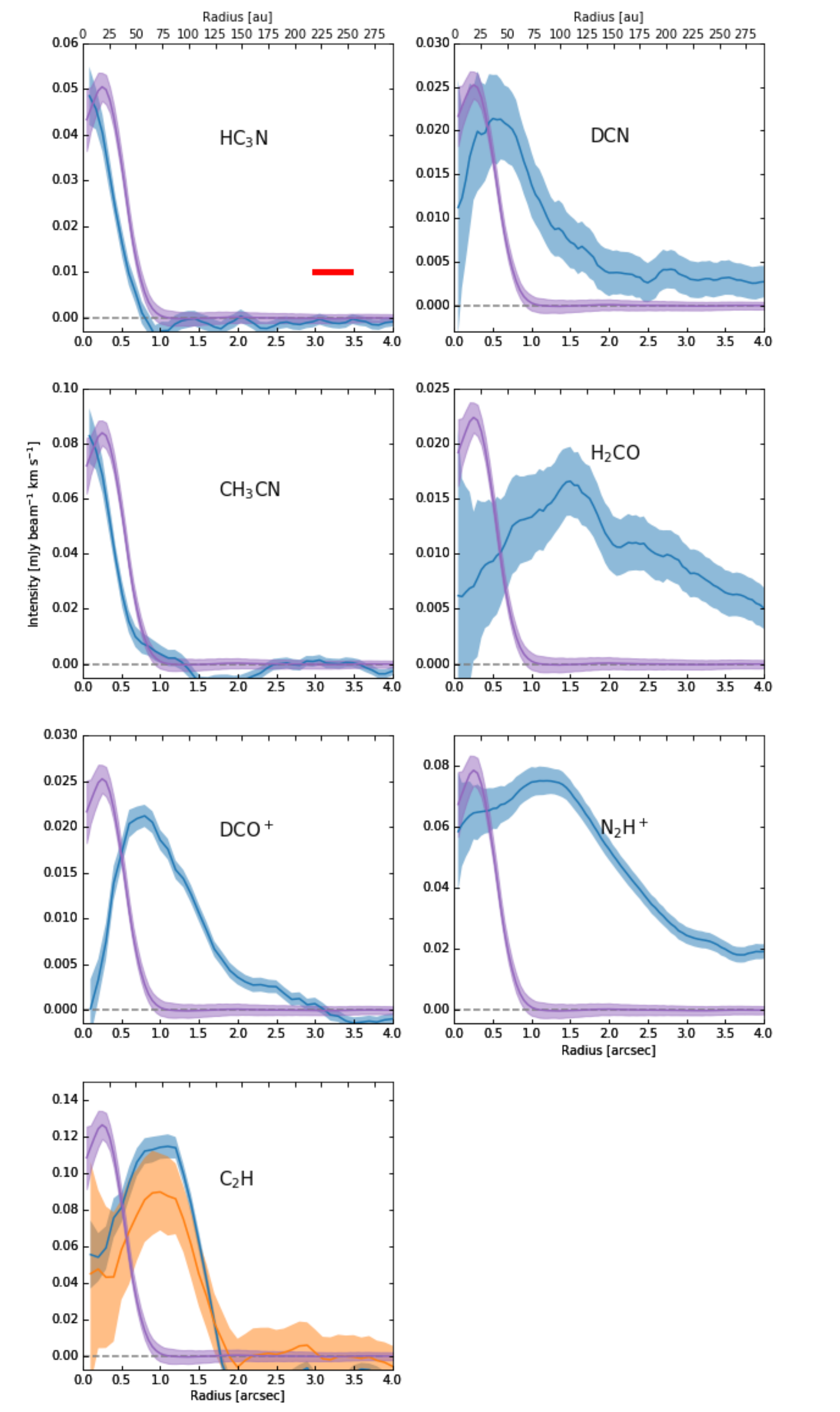}
\caption{Radial profiles extracted from integrated intensity emission-line images of molecular species displaying relatively sharp ring-like morphologies  (left panels) and more diffuse ring-like morphologies  (right panels). The blue and orange curves in the lower left panel represent the radial profiles of the integrated intensities of the 262.004 GHz and 262.064 GHz hyperfine complexes of C$_2$H, respectively.
The appropriately normalized radial profile of 264 GHz continuum emission is overlaid (as a purple curve) in each panel, for reference. The red bar in the upper left panel indicates a representative beamwidth of $0.6''$.
}
\label{fig:RingsRadialProfiles}
\end{center}
\end{figure}

The molecular emission line images in Figs.~\ref{fig:CPImages6by6} and \ref{fig:RingImages6by6} are grouped according to their morphological properties \citep[see also][]{Huang2017}, 
wherein Fig.~\ref{fig:CPImages6by6} displayes images with centrally peaked morphologies and Fig.~\ref{fig:RingImages6by6} displays images with ring-like morphologies that range from distinct to somewhat diffuse. Correspondingly, in Figs.~\ref{fig:CPradialProfiles} and \ref{fig:RingsRadialProfiles}, respectively, we display radial profiles extracted from the centrally-peaked and ring-like molecular emission line images (see \S 3.2).
\begin{description}
\item[Centrally peaked morphologies (Figs.~\ref{fig:CPImages6by6}, \ref{fig:CPradialProfiles}).] The images of C and O isotopologues of CO and C and N isotopologues of HCN all display centrally peaked morphologies. The line emission from the rare CO and HCN isotopologues is far more compact than that of the most abundant isotopologues ($^{12}$C$^{16}$O and H$^{12}$C$^{14}$N), as expected in light of the smaller optical depths and resulting lower fluxes in the lines of the rare isotopologues. 
\item[Distinct ring-like morphologies (Figs.~\ref{fig:RingImages6by6}, \ref{fig:RingsRadialProfiles}).] The HC$_3$N, CH$_3$CN, DCO$^+$, and C$_2$H line images display ring-like emission morphologies wherein the emission rings have clearly defined central holes, relatively sharp outer edges, or both. The emission from the cyanide group molecules, HC$_3$N and CH$_3$CN, arises from compact regions whose outer radii are similar to those of the 1.1--1.4 mm continuum emission ring (see \S~\ref{sec:radialProfiles}). Due to their low signal-to-noise ratios and small angular sizes, we find that the depths of the inner holes within the HC$_3$N and CH$_3$CN emitting regions are sensitive to the adopted visibility weighting during the image reconstruction (cleaning) process, while the degree of azimuthal asymmetry is sensitive to channel noise clipping in constructing the moment 0 images. However, the overall ring-like morphologies of HC$_3$N and CH$_3$CN --- with the CH$_3$CN ring possibly more asymmetric and filled in than HC$_3$N --- appear to be a robust result of these observations. The size scales of the DCO$^+$ and C$_2$H rings are much larger than that of the cyanide group molecules, with integrated line intensities that peak far outside the continuum emission ring. The C$_2$H ring exhibits sharper edges than that of DCO$^+$ and, as is seen more clearly in radial profiles extracted from these images (\S~\ref{sec:radialProfiles}), the C$_2$H also peaks at a larger radius.
\item[Diffuse ring-like morphologies (Figs.~\ref{fig:RingImages6by6}, \ref{fig:RingsRadialProfiles}).] The DCN, H$^{13}$CO$^+$,  N$_2$H$^+$, and H$_2$CO line images all display diffuse emission morphologies and appear as rings to a greater or lesser degree. The DCN(4--3) image presented here, which has higher signal-to-noise ratio than the DCN(3--2) image presented in \citet{Huang2017}, confirms the ring-like DCN emission morphology hinted at in that survey. Evidently, the diffuse DCN ring is more compact than the sharper C$_2$H or DCO$^+$ rings, while the H$^{13}$CO$^+$ ring dimensions appear to more closely resemble those of DCN than those of DCO$^+$. The radius of peak integrated line intensity within the bright N$_2$H$^+$ ring is comparable to the radius of peak integrated C$_2$H line intensity but, with its filled-in central hole and outer halo, the N$_2$H$^+$ ring is much larger and more diffuse in appearance. The outer radius of the H$_2$CO emitting region appears to be similar to that of CO, but (unlike CO) the H$_2$CO has a central hole. As in the case of DCO$^+$ and C$_2$H, the inner holes of the N$_2$H$^+$ and H$_2$CO  line emission rings, though poorly defined, appear similar to the outer radii of the continuum rings.
\end{description}

\subsection{Radial Profiles of Integrated Molecular Line Intensity \label{sec:radialProfiles}}

In Figs.~\ref{fig:CPradialProfiles} and \ref{fig:RingsRadialProfiles}, we display radial profiles extracted from unclipped versions of the moment 0 molecular emission line images of V4046 Sgr, as well as the radial profile of the 264 GHz continuum image. These profiles were extracted after deprojecting the images assuming a disk inclination $i=33.5^\circ$ and position angle of 67$^\circ$ \citep{Rosenfeld2013}. The error bars indicate statistical uncertainties, which we estimated as $\sigma_r = \sigma_{xy} \, (\sqrt{\Delta V / \delta v}) \, (1/\sqrt{N}) $, where $\sigma_{xy}$ is the rms flux density deviation in a channel map (as determined from an emission-free region of the map), 
$\Delta V$ is the linewidth, $\delta v$ is the channel width (such that $\Delta V / \delta v$ is the number of velocity channels over which the line emission was integrated), and $N$ is the number of pixels included within the radial bin. 

Fig.~\ref{fig:CPradialProfiles} provides a comparison of the radial profiles of integrated emission from isotopologues of CO and HCN, which have centrally-peaked image morphologies (Fig.~\ref{fig:CPImages6by6}). Some caution must be applied in comparing these profiles in detail, since the beamwidths of the CO isotopologue images are $\sim$40\% larger than those of the HCN isotopologues (Table~\ref{tbl:LibraryImages}). The radial profiles of both $^{12}$CO and H$^{12}$CN show inflection points (slope changes) at $\sim$1.5$''$ ($\sim$100 AU) that are suggestive of bright cores surrounded by larger halo structures.  Such features can be produced by subtraction of optically thick continuum emission from optically thick line emission \citep[][]{Weaver2018}, but that is unlikely to be the case here, given that the inflection points lie well beyond the outer edge of the continuum ring ($\sim$0.7$''$).
The $^{12}$CO falls off steeply with radius within the inner continuum ring, whereas the H$^{12}$CN profile is shallow within this same region. The radial profiles demonstrate that detectable emission extends to $\sim$2.5$''$ ($\sim$180 AU) and $\sim$4$''$ ($\sim$300 AU) for H$^{12}$CN and $^{12}$CO, respectively, with $^{12}$CO intensity displaying a sharp outer edge that is best seen in the comparison of radial intensity profiles of CO isotopologue emission presented in \S 4.1. 

\begin{figure}[htbp]
\begin{center}
\includegraphics[width=4.0in]{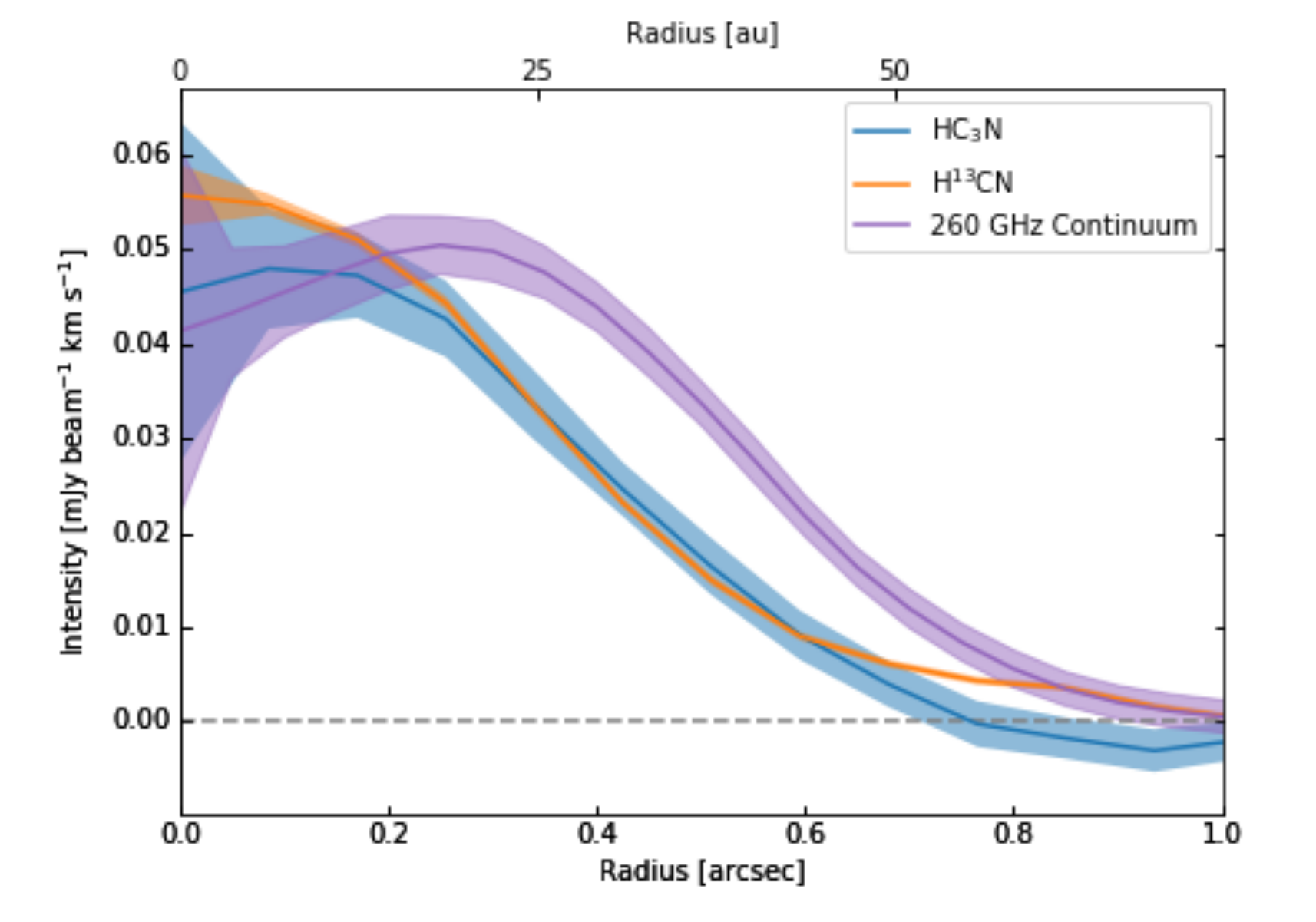}
\caption{Comparison of the radial intensity profile of HC$_3$N with those of H$^{13}$CN and continuum emission within the inner 1$''$ of the V4046 Sgr disk. The H$^{13}$CN and continuum profiles have been rescaled, for clarity.
}
\label{fig:HC3NvsH13CNcomp}
\end{center}
\end{figure}

In the left and right panels of Fig.~\ref{fig:RingsRadialProfiles}, we compare the radial profiles of integrated emission from molecular lines exhibiting sharply ring-like and more diffuse ring-like morphologies, respectively (Fig.~\ref{fig:RingImages6by6}). Such a direct comparison of these radial profiles is enabled by the similar beam sizes of the data collected for all of these transitions (i.e., their beamwidths differ by $<$15\%; Table~\ref{tbl:LibraryImages}). A sequence is apparent in the radial position of peak intensity and (hence) central hole size within these rings of molecular line emission. The smallest rings are those of the cyanide group molecules (CH$_3$CN and HC$_3$N), which peak near $\sim$0.25$''$ ($\sim$18 AU), just inside the peak continuum emission (see Fig.~\ref{fig:HC3NvsH13CNcomp}). The largest rings are the sharp-edged C$_2$H and diffuse N$_2$H$^+$, both of which peak at $\sim$1.2$''$ ($\sim$90 AU), and H$_2$CO, which displays a broad peak at $\sim$1.6$''$ ($\sim$115 AU). The emission rings of the deuterated species DCN and DCO$^+$, which peak at $\sim$0.6$''$ ($\sim$45 AU) and $\sim$0.8$''$ ($\sim$60 AU), respectively, are intermediate in size. While we do not display the radial profile of H$^{13}$CO$^+$ due to its relatively poor S/N ratio, this faint ring appears to more closely trace DCN than DCO$^+$ \citep[Fig.~\ref{fig:RingImages6by6}; see also Fig.~2 in][]{Huang2017}.

The upper left panels of Fig.~\ref{fig:RingsRadialProfiles} well illustrate the close correspondence of the radial profiles of emission from the cyanide group molecules (HC$_3$N and CH$_3$CN) to each other and to that of the 264 GHz continuum continuum emission profile, which peaks at $\sim$0.3$''$ ($\sim$22 AU) and has an outer radius (at 10\% of peak) of $\sim$0.8$''$ ($\sim$60 AU). The peak intensities of HC$_3$N and CH$_3$CN emission appear to lie just inside that of the continuum, and the lack of a central hole is apparent in the case of CH$_3$CN. The radial distributions of HC$_3$N and CH$_3$CN line emission both display sharp outer radial cutoffs, falling to $\sim$10\% of peak intensity at $\sim$0.7$''$ ($\sim$50 AU). 

\subsection{Line Profiles}

\begin{figure}[htbp]
\begin{center}
\includegraphics[width=3in]{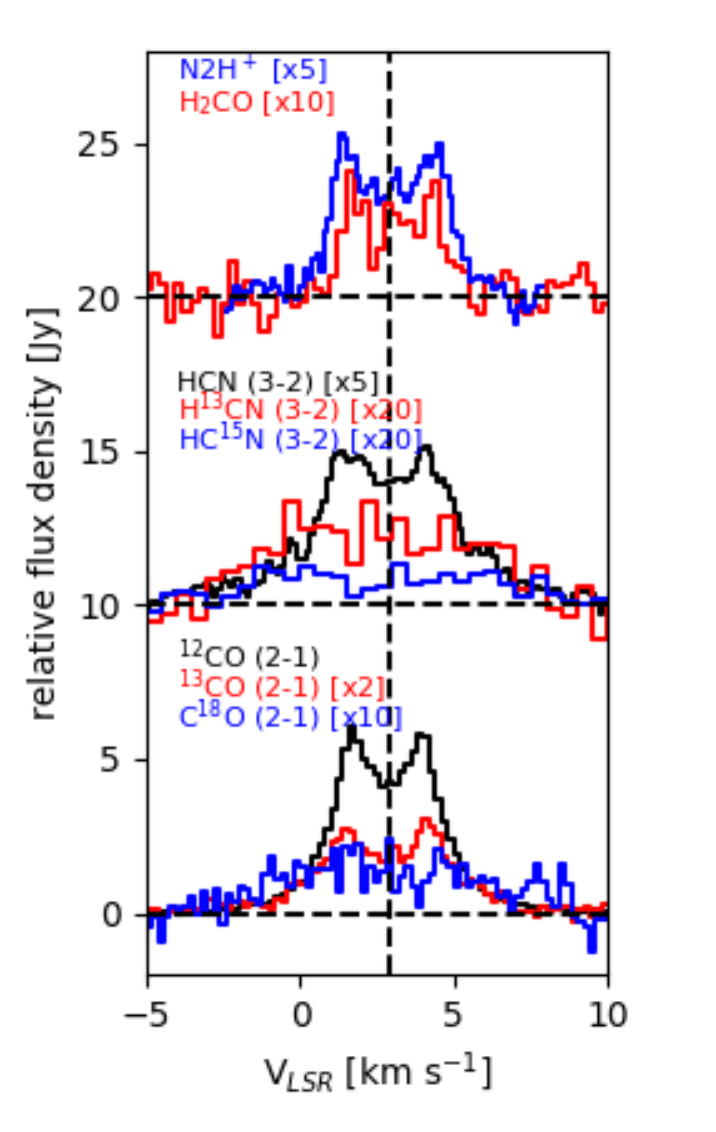}
\includegraphics[width=3in]{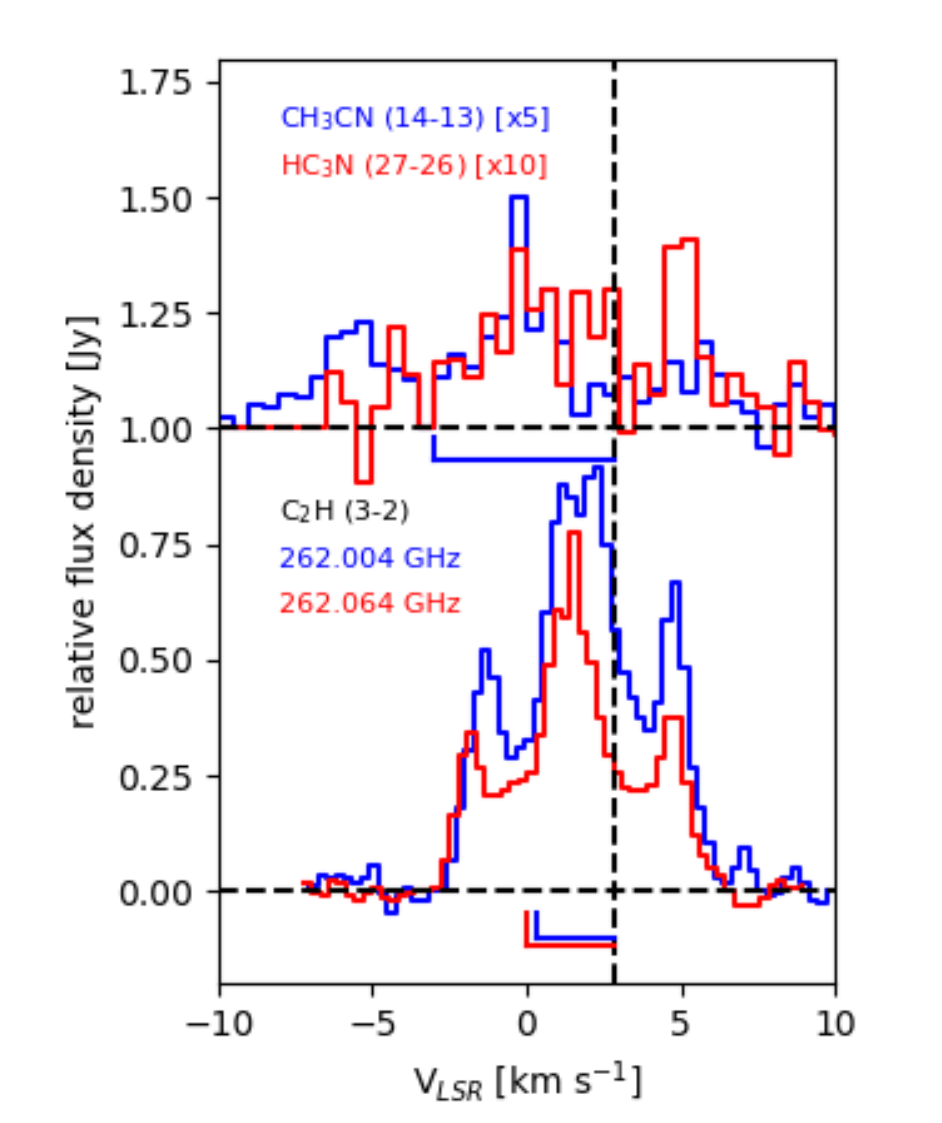}
\caption{Line profiles extracted from the emission line image cubes. In each panel, the vertical dashed lines indicate the systemic velocity of V4046 Sgr (2.8 km s$^{-1}$) and the horizontal dashed lines mark the spectral baselines. Left panel: line profiles of isotopologues of CO (bottom) and HCN (middle), and profiles of N$_2$H$^+$ and H$_2$CO (top). Right panel: line profiles of the two brightest hyperfine complexes of C$_2$H (bottom) and the complex nitriles (HC$_3$N and CH$_3$CN; top), with the velocity offsets corresponding to the hyperfine splitting of the CH$_3$CN and C$_2$H transitions indicated.}
\label{fig:LineProfiles}
\end{center}
\end{figure}

In Fig.~\ref{fig:LineProfiles}, we display line profiles extracted from the interferometric data cubes for molecular species observed by ALMA subsequent to the observations presented in \citet{Huang2017} and \citet{Guzman2017} or whose profiles were not presented in those papers. The extraction regions were ellipses whose major and minor axes approximately correspond to the $\sim$3$\sigma$ noise levels in the individual velocity channel images. Whereas the $^{12}$CO and $^{13}$CO line profiles display the classical double-peaked profiles characteristic of Keplerian rotation \citep[as was already apparent in single-dish observations;][]{Kastner2008}, the profile of C$^{18}$O --- which is presented here for the V4046 Sgr disk for the first time --- appears more rounded, with stronger wings relative to the line core. This difference reflects the smaller detectable extent of the C$^{18}$O emission, which effectively supresses the emission at the low radial velocities characteristic of the outer disk relative to the higher-velocity emission characteristic of the central disk. Similar arguments appear to pertain to the line profiles of isotopologues of HCN, wherein the line profile of the most abundant isotopologue appears double-peaked and those of the rare isotopologues more rounded. The line profiles of N$_2$H$^+$ and H$_2$CO lack high-velocity wings, as expected given their large, diffuse, ringlike morphologies. The triple-peaked C$_2$H and CH$_3$CN line profiles result from the superposition of double-peaked emission lines in neighboring hyperfine transitions \citep[compare with C$_2$H and CH$_3$CN line profiles displayed in][respectively]{Kastner2014,Oberg2015}.

\section{Discussion}

\subsection{Radial dependencies of CO isotopologue emission ratios}

Given their similar beam areas (which differ by $<$5\%) and comparable sensitivities,
the images in $J = 2\rightarrow1$ emission from $^{12}$CO, $^{13}$CO, and C$^{18}$O
(Fig.~\ref{fig:CPImages6by6}) afford the opportunity to investigate the ratios of CO isotopologue emission line intensity as functions of radial position across the V4046 Sgr disk. These ratios in turn can be used to assess the optical depths of $^{12}$CO, $^{13}$CO, and C$^{18}$O emission. Specifically, given an assumed (constant) isotopologue abundance ratio $X_{1,2}$ within the disk, the ratio of line emission intensities in two isotopologues at radial position $r$ can be related to optical depth in the first (more abundant) isotopologue, $\tau_1$, via  
$$
R_{1,2}(r) = \frac{1-\exp{(-\tau_1(r))}}{1-\exp{-(\tau_1(r)/X_{1,2})}}
$$
\citep[e.g.,][carried out such an analysis of $\tau (r)$ for the TW Hya disk]{Kastner2014,Schwarz2016}. 
Caution must be applied in interpreting estimates for $\tau_1(r)$ as deduced from $R_{1,2}(r)$, given that emission from the different isotopologues likely arises from different vertical disk layers \citep[e.g.,][]{ZhangBergin2017}. Furthermore, the optical depth can vary significantly both spatially (within the synthesized beam) as well as across the line profile over which the intensity is integrated. Hence, we restrict the present discussion to a qualitative analysis of the radial regimes over which the various isotopologue emission lines appear to be optically thick or thin.

\begin{figure}[htbp]
\begin{center}
\includegraphics[width=3in]{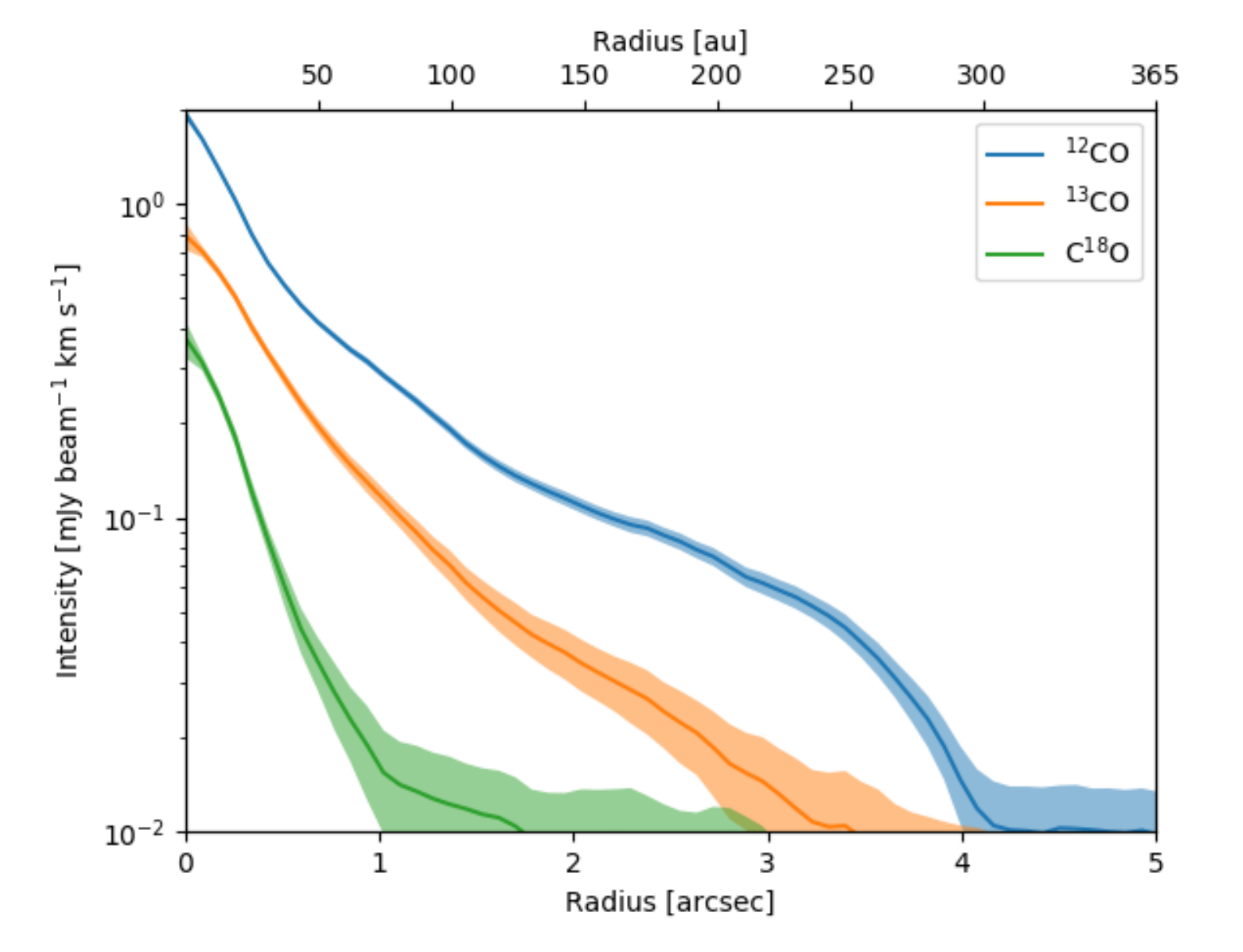}
\includegraphics[width=3in]{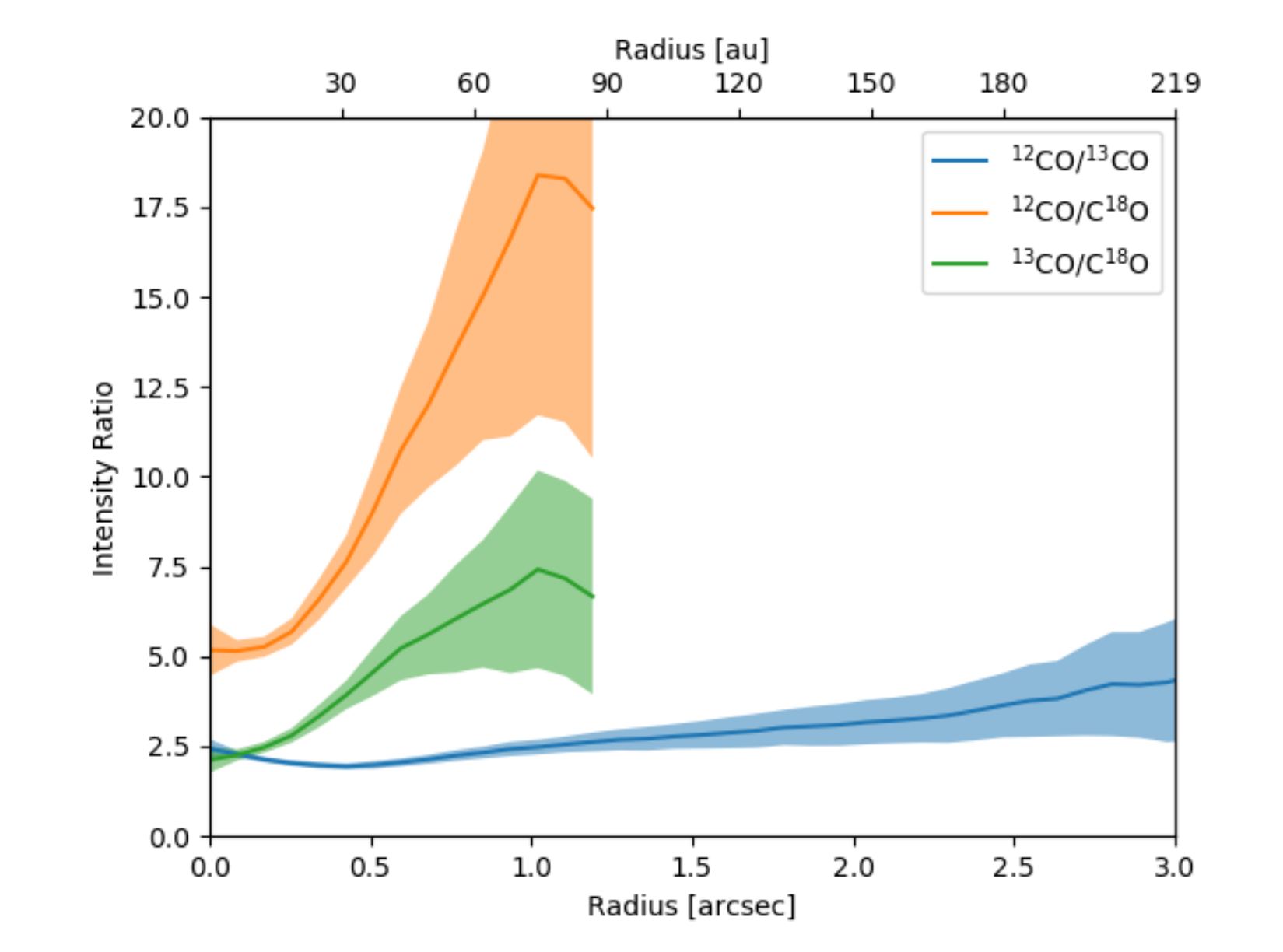}
\caption{Left: radial profiles of integrated intensity of CO
isotopologue line emission. Right: radial profiles of ratios of CO isotopologue emission line intensity.}
\label{fig:RadProfsCO}
\end{center}
\end{figure}

Comparisons of the radial profiles in the various isotopologues, as well as the radial profiles of the intensity ratios of CO isotopologue emission, are presented in Fig.~\ref{fig:RadProfsCO}. The integrated intensity of $^{12}$CO emission shows a precipitous drop at $\sim$4$''$ ($\sim300$ AU), indicative of either the steep outer edge of the gas disk or a sharp decline in CO abundance; the latter would presumably be due to UV photodissociation. The maximum radii of detectable  $^{13}$CO and C$^{18}$O emission are $\sim$3$''$ ($\sim$220 AU) and $\sim$1$''$ ($\sim$75 AU), respectively. These radii set the respective limits within we which we can measure the $^{12}$CO:$^{13}$CO and $^{13}$CO:C$^{18}$O (and $^{12}$CO:C$^{18}$O) emission line ratios. We find the ratio of $^{12}$CO:$^{13}$CO line emission intensities to be essentially independent of $r$ given the uncertainties, with a value $R_{12,13} \sim 2.5$, while the $^{12}$CO:C$^{18}$O line ratio increases from $R_{12,18} \sim 5$ to $R_{12,18} \sim 18$ over the radial range $r \stackrel{<}{\sim} 25$ AU to $r \sim 75$ AU. The ratio of $^{13}$CO:C$^{18}$O line emission intensities $R_{13,18}$ increases from $R_{13,18} \sim 2$ to $R_{13,18} \sim 7$ over this same radial range. 

For the values of $X_{12,13}$ and $X_{12,18}$ typically assumed in the astrophysical literature --- i.e., $X_{12,13}$ from $\sim$40 to $\sim$70, and $X_{12,18} \sim 480$ \citep[][and references therein]{Kastner2014,ZhangBergin2017} --- these small values of $R_{12,13}$ and $R_{12,18}$ imply, not surprisingly, that the $^{12}$CO(2--1) emission is very optically thick (i.e., $\tau_{12} >> 1$) across the disk surface. Meanwhile, the observation that $R_{13,18} < 7$ (i.e., $R_{13,18} <  X_{13,18}$, assuming $X_{12,13} = 70$) within $\sim$75 AU indicates that $\tau_{13}> 1$, at least in the inner disk; the small, near-constant value of  $R_{12,13}$ furthermore implies that $^{13}$CO(2--1) is optically thick throughout the disk. 
Assuming that the ratio of $^{13}$CO to C$^{18}$O optical depths is identical to their abundance ratio, the foregoing results for $\tau_{13}$ for V4046 Sgr also imply that C$^{18}$O is optically thin throughout much of the disk. These results for  $\tau_{13}$ and $\tau_{18}$ are similar to those inferred for TW Hya by \citet{Schwarz2016}. We note, however, that \citet{ZhangBergin2017} find that C$^{18}$O(3--2) is optically thick in the innermost  regions ($r \le 20$ AU) of the TW Hya disk. 

If both $^{12}$CO(2--1) and $^{13}$CO(2--1) are optically thick, their line ratio is diagnostic of the characteristic temperatures of the regions from which the bulk of the emission in each line originates. Hence, Fig.~\ref{fig:RadProfsCO}  indicates that the optically thicker $^{12}$CO(2--1) emission arises from a warmer and hence higher-lying disk layer than $^{13}$CO(2--1), as expected given the vertical temperature inversion that is a feature of models of irradiated disks \citep[e.g.,][]{ZhangBergin2017}. Higher spatial resolution ALMA imaging of CO isotopologue emission from the V4046 Sgr disk, as well as observations of $^{13}$C$^{18}$O emission analogous to those carried out for TW Hya \citep{ZhangBergin2017}, would test the foregoing qualitative inferences concerning the radial dependencies of the optical depths of CO isotopologue emission and would provide important additional constraints on disk vertical structure, molecular mass, and $^{12}$C:$^{13}$C abundance ratio.

\subsection{The potential connection between complex nitriles and dust}

V4046 Sgr stands out among disks surveyed thus far by ALMA for its unusually bright emission from both HC$_3$N and CH$_3$CN \citep{Bergner2018}. As in the case of MWC 480, the first disk detected in both of these complex nitrile (cyanide-bearing) species \citep{Oberg2015}, there appears to be a spatial correspondence between the HC$_3$N and CH$_3$CN line emission and the continuum emission from large (mm-sized) dust grains in the V4046 Sgr disk. However, the subarcsecond ALMA imaging of V4046 Sgr presented here makes apparent the ring-like morphologies of HC$_3$N and, possibly, CH$_3$CN (Fig.~\ref{fig:RingImages6by6}). These morphologies closely correspond to those of the continuum emission as well as scattered light from small grains just interior to the continuum ring \citep{Rapson2015a} --- and they contrast with the centrally peaked emission from the rare HCN isotopologues (Fig.~\ref{fig:HC3NvsH13CNcomp}) --- suggesting a connection between the presence of dust rings and the production of gas-phase HC$_3$N and CH$_3$CN.  Various pure gas-phase production routes for HC$_3$N and CH$_3$CN and potential grain surface chemistry production of CH$_3$CN are discussed by \citet{Bergner2018} in the context of their survey of complex nitriles in V4046 Sgr and several other disks. Here, we point out one possible interpretation of the apparent morphological correspondence between emission from these species and that of emission from (and scattering off) dust grains within the V4046 Sgr disk.

Specifically, based on the results of simulations \citep{2014A&A...563A..33W,Cleeves2016} and laboratory experiments \citep{2013MNRAS.433.3440M}, it is possible that the ring-like HC$_3$N and CH$_3$CN emitting regions are generated by stellar irradiation of ice-coated grains. In such a scenario, HC$_3$N, CH$_3$CN, and other, more complex organics either form in the disk, or were inherited from the prestellar core that spawned V4046 Sgr. In either case, a significant mass of organics likely subsequently accretes (freezes out) onto grains in the present-day disk and/or at earlier phases of the protostellar (Class 0/I) disk. The intense stellar UV and X-irradiation from V4046 Sgr \citep[e.g.,][]{Argiroffi2012} impinging on the well-defined rings of dust grains within the disk then results in efficient desorption of the organic ice coatings, generating detectable column densities of HC$_3$N and CH$_3$CN \citep[][]{2014A&A...563A..33W}. Interestingly, HC$_3$NH$^+$ is one of the dominant products of X-irradiation of pyrimidine ice-coated grains in laboratory experiments \citep{2013MNRAS.433.3440M}. This suggests that the abundances of HC$_3$N might be enhanced in UV- and X-irradiated regions that are sufficiently rich in pyrimidine and (perhaps) other organic ices, such that both the abundance of HC$_3$NH$^+$ and the molecular gas ionization fraction are elevated. On the other hand, it is not clear whether either HC$_3$N or CH$_3$CN would then survive long in the gas phase, given the large photodissociation rates within such a hostile radiation environment.

Similar arguments pertain to the production of gas-phase H$_2$CO via UV and X-ray photodesorption of hydrogenated grain (CO) ice mantles \citep{2014A&A...563A..33W,Oberg2017}. Indeed, the \citet{2014A&A...563A..33W} modeling predicts that the column density of gas-phase H$_2$CO so generated should peak at, and remain elevated out to, much larger disk radii than the column densities of HC$_3$N and CH$_3$CN. Qualitatively, these model predictions also seem to be borne out by the ALMA imaging presented here (Fig.~\ref{fig:RingsRadialProfiles}; see also \S \ref{sec:Rings}).  We note, however, that the \citet{2014A&A...563A..33W} model assumes a smoothly varying disk dust component, which obviously does not well describe evolved protoplanetary disks like the one orbiting V4046 Sgr, with its relatively narrow, compact ring of large grains and interior small-grain dust ring/gap system. The HC$_3$N and CH$_3$CN images presented here hence serve as motivation for future simulations aimed at ascertaining whether and how the processes of photodesorption and photodissociation of organics in a protoplanetary disk depend on the extent of dust grain coagulation and transport within the disk. Given that thermal excitation of the $\sim$260 GHz transitions observed here requires gas kinetic temperatures of $\sim$100 K, sensitive observations of lower-lying transitions of HC$_3$N and CH$_3$CN in V4046 Sgr would also help test the hypothesis that photodesorption is responsible for production of these molecules. As noted by \citet{Bergner2018}, the extant data is not sensitive to emission from these molecules within cooler disk regions at larger radii.

\subsection{The nested molecular rings orbiting V4046 Sgr \label{sec:Rings}}

Beyond the $\sim$50 AU outer radius of the continuum and  HC$_3$N and CH$_3$CN emission in the V4046 Sgr disk, emission from the species DCN, DCO$^+$, N$_2$H$^+$, C$_2$H, and H$_2$CO appears as a sequence of rings with increasing radius of peak intensity (i.e., intensity peaks at $\sim$45, 60, 90, 90, and 110 AU, respectively) and varying sharpness in terms of inner and outer cutoff radii (Fig.~\ref{fig:RingsRadialProfiles}). Among these rings, the sharp-edged morphology of C$_2$H is perhaps most striking; this emission ring is modeled and discussed in detail in \S~\ref{sec:C2H}. Here, we briefly comment on some potential implications of this molecular ring ``sequence.''

The main routes proposed for formation of DCO$^+$ and DCN require the presence (destruction) of H$_2$D$^+$ and CH$_2$D$^+$, respectively \citep[][and references therein]{Huang2017}. Because survival of H$_2$D$^+$ requires temperatures $\stackrel{<}{\sim}$30 K, whereas survival of CH$_2$D$^+$ is energetically favorable up to $\sim$80 K, one expects DCN to peak in abundance at smaller disk radii than DCO$^+$ \citep{Huang2017}. Such a relationship is indeed observed in the case of V4046 Sgr. As formation of DCO$^+$ also depends on the presence of gas-phase CO, one further expects DCO$^+$ to be confined to a disk temperature regime and, hence, (midplane) radial regime where the disk is not warm enough for destruction of H$_2$D$^+$ (which requires temperatures in excess of $\sim$30 K) but not cold enough that there is significant freezeout of CO (temperatures below $\sim$25 K). This rather narrow range of temperatures favorable to DCO$^+$ formation may explain the ring-like appearance of DCO$^+$ in V4046 Sgr and other disks \citep[e.g., HD 163296;][]{Qi2015}. On the other hand, \citet{Huang2017} caution against concluding that DCO$^+$ serves to trace the specific disk regime that lies just above the CO freezeout temperature (snow line), citing the lack of a clear relationship between the radial extents of DCN and DCO$^+$ in many of the disks they surveyed. 

As discussed in \citet{Qi2013a,Qi2013,Qi2015} and \citet{Oberg2017}, mm-wave emission lines of the species N$_2$H$^+$ and H$_2$CO (particularly the former) should provide particularly effective tracers of the CO snow line, albeit for complementary reasons: N$_2$H$^+$ is destroyed via reactions with gas-phase CO, forming HCO$^+$; while, as noted, H$_2$CO can be efficiently generated through hydrogenation of CO ice mantles on dust grains followed by photodesorption. Hence, one expects the largest N$_2$H$^+$ and H$_2$CO abundance enhancements in disk regions cold enough for CO freezeout \citep[although there are also warm formation pathways for H$_2$CO;][]{Oberg2017}. In general terms, this expectation appears to be borne out in V4046 Sgr, i.e., the N$_2$H$^+$ and H$_2$CO rings both reach their peak intensities beyond the peaks of the DCN and DCO$^+$ rings. 
However, in contrast to the TW Hya and HD 163296 disks --- both of which have N$_2$H$^+$ emission rings with sharp inner cutoffs \citep[][]{Qi2013,Qi2015} --- there is no clear central hole within the N$_2$H$^+$ ring in the V4046 Sgr disk. As a result, the precise radial position of putative midplane CO freezeout is difficult to ascertain for V4046 Sgr on the basis of its radial profile of N$_2$H$^+$ line intensity. Such an interpretation is rendered even more difficult by the possibility that the N$_2$H$^+$ column density may smoothly increase beyond the midplane CO snow line, and/or that there may be significant abundances of N$_2$H$^+$ in disk surface layers \citep[see discussions in][]{Nomura2016,vantHoff2017}.  Further, detailed analysis of the ALMA N$_2$H$^+$ and H$_2$CO imaging results for V4046 Sgr will be presented in a forthcoming paper (Qi et al., in prep.).  

\subsection{An empirical model for the C$_2$H ring \label{sec:C2H}}

Our molecular line surveys of TW Hya, V4046 Sgr, and the disk orbiting LkCa 15 (age $\sim$5 Myr) established that the emission line intensities of C$_2$H from these evolved disks rival or exceed those of, e.g., $^{13}$CO \citep{Kastner2014,Punzi2015}. Followup SMA and ALMA imaging of C$_2$H emission from TW Hya \citep{Kastner2015,Bergin2016} revealed that the C$_2$H  line emission exhibits a ring-like morphology. Our ALMA imaging has now established that the C$_2$H emission from V4046 Sgr displays the same well-defined, ring-like morphology \citep[a result already apparent in our previous SMA C$_2$H imaging;][]{Kastner2016IAUS}, indicating that sharp-edged C$_2$H rings are a common feature of evolved disks. 

Given the paucity of bright molecular line tracers of disk chemical and physical conditions, it is essential to understand the production mechanism(s) responsible for the large abundances of C$_2$H and its ring-like distribution in these and other disks \citep[e.g., DM Tau;][]{Bergin2016}. 
Based on the SMA C$_2$H imaging results for TW Hya and consideration of the excitation of C$_2$H, we proposed that the C$_2$H ring traces particularly efficient UV/X-ray photo-destruction of hydrocarbons derived from small grains and grain ice mantles in the low-density ($n \ll 10^7$ cm$^{-3}$), large-grain-depleted surface layers of the outer ($>$45 AU) regions of the TW Hya disk \citep{Kastner2015}. 
Subsequent ALMA C$_2$H imaging and accompanying modeling supports the general notion that efficient C$_2$H production in evolved disks is a signpost of grain size segregation and stellar irradiation \citep{Bergin2016}. However, \citet{Bergin2016} concluded that TW Hya's ring-like C$_2$H emission morphology is also a result of C depletion in the inner disk, and that the emission arises from disk layers with $n > 10^6$ cm$^{-3}$. If so, then the presence of a C$_2$H ring would most likely be the result of pure gas-phase and/or gas-grain processes deep within the disk, with little or no influence from stellar irradiation \citep[see discussion in][and refs.\ therein]{Kastner2015}.

The foregoing uncertainty in the vertical location of C$_2$H within the TW Hya disk is a consequence of the model degeneracies inherent to its near-pole-on orientation \citep[$i=7^\circ$;][and refs.\ therein]{Andrews2012}.  Given the more intermediate inclination of the V4046 Sgr disk \citep[$i=33.5^\circ$;][]{Rosenfeld2013}, this disk potentially provides a means to discriminate between the various alternative C$_2$H production models --- i.e., surface-layer, irradiation-driven production vs.\ midplane, pure gas-phase production --- to explain the presence of bright C$_2$H rings in these two disks. 

To investigate the potential of the ALMA C$_2$H imaging in this regard, we have modeled the ring-like C$_2$H  line emission following the methodology described in detail in \citet{Qi2013} and \citet{Kastner2015}. Briefly, as in \citet{Qi2013}, we adopt a physically self-consistent accretion disk model \citep[][and references therein]{DAlessio2006} that matches the spectral energy distribution (SED), with the disk geometric (i.e., scale height and surface density profile) parameters fixed to values previously determined via SED fitting for V4046 Sgr \citep{Rosenfeld2013}. Within this framework, we then investigate a limited parameter set that describes the disk layer from which the molecular emission originates. Specifically,  we invoke the presence of a molecular emission layer of constant abundance that is confined to a range of vertical column densities between $10^{Z_1} \times (1.59\times10^{21})$ cm$^{-2}$ and $10^{Z_2} \times (1.59\times10^{21})$ cm$^{-2}$. 
The column density of the molecular layer is then allowed to vary within radial annuli ranging from an arbitrarily small inner radius $R_0$ that is much smaller than the  beam size (in this case, $R_{0} = 10$ AU), through an effective inner radius $R_{in}$ (where the column density increases to detectable values; see below), out to an outer cutoff radius $R_{out}$. The model grid interval is 10 AU in the outer disk ($r>50$ AU) to roughly 5 AU in the inner disk ($r<50$ AU). The radial column density dependence is characterized by a set of radial power-law slopes $p_n$ at break points $R_{n-1}$; for present purposes we set $n=2$. For a selected set of free parameters of interest --- in this case, $R_{in}, R_{out}, p_1, p_2, R_1$, for a given $Z_1, Z_2$ --- radiative transfer calculations are carried out via the \verb+RATRAN+ code to determine the resulting sky-projected integrated line intensity distribution. Model integrated line intensity distributions are then realized over a wide range for each parameter, and fitting to the interferometric molecular emission line data is performed in visibility space via a grid search approach. 

\begin{figure}[htbp]
\begin{center}
\includegraphics[width=3.1in]{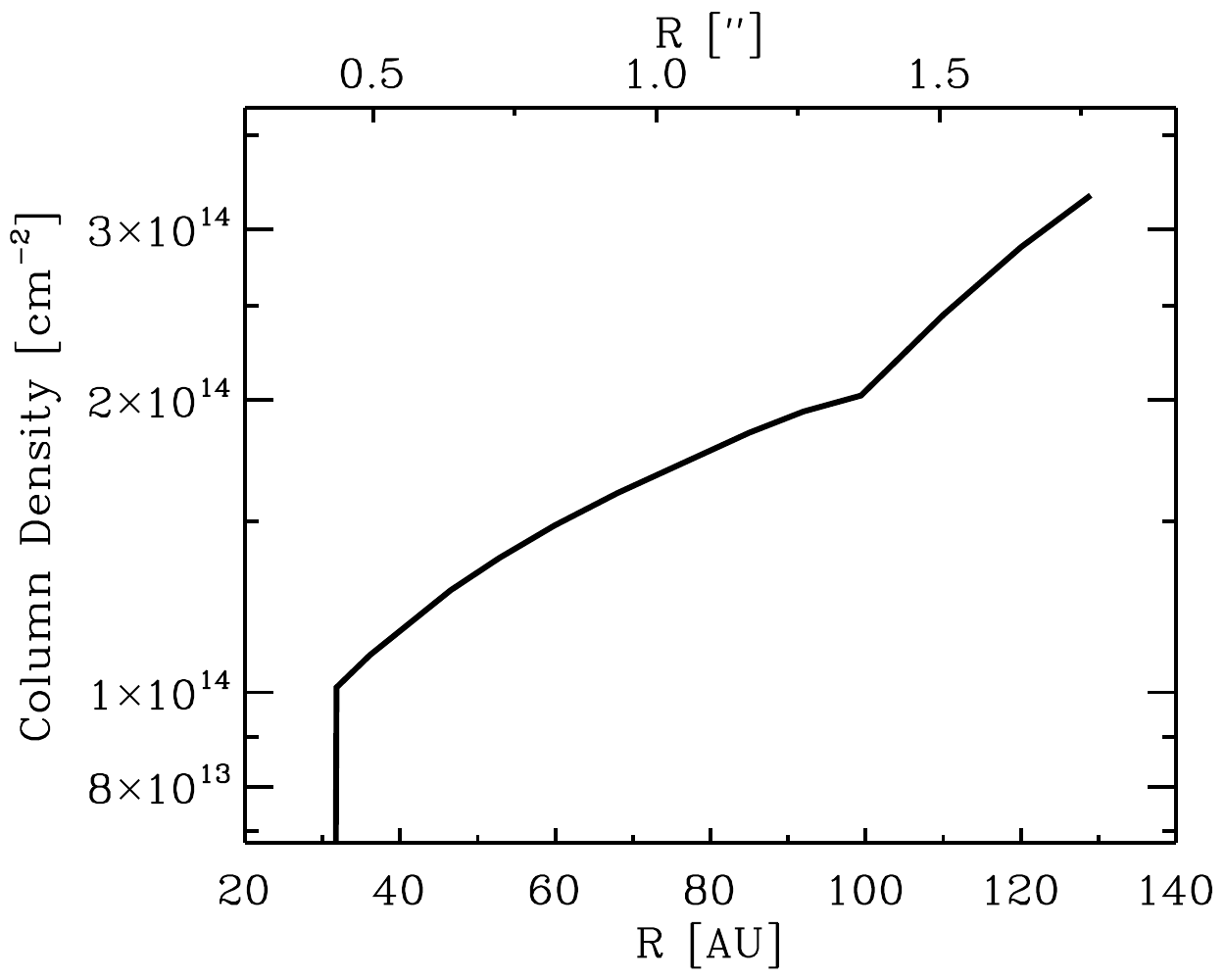}
\includegraphics[width=2.9in]{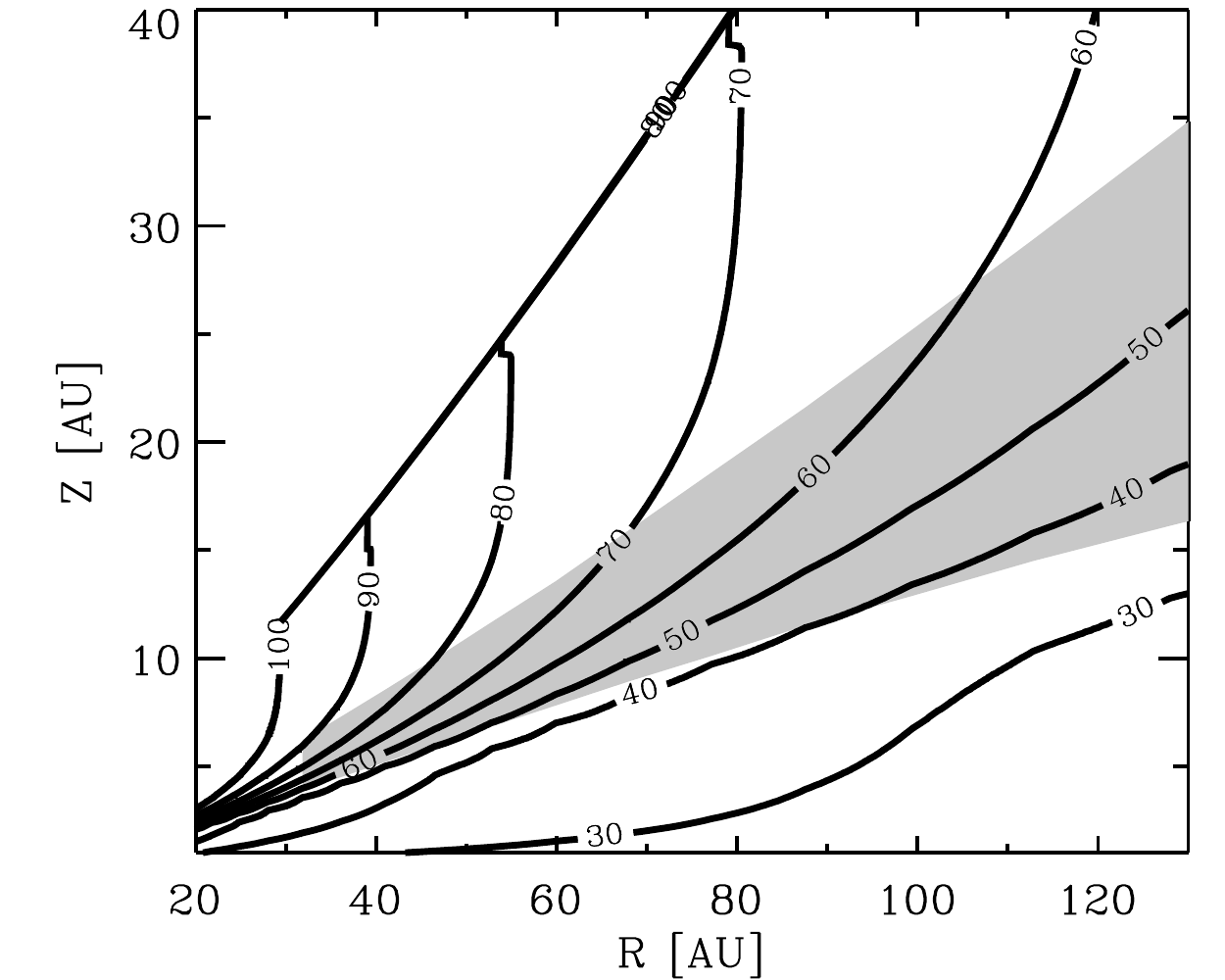}
\caption{Left: Radial profile of C$_2$H column density, for the best-fit model. Right: Radial and vertical distribution of the C$_2$H emitting layer in the best-fit model  (grey shaded region), overlaid on contours of gas temperature. 
}
\label{fig:C2Hmodel}
\end{center}
\end{figure}

We carried out this model-fitting procedure for the 262.004 GHz hyperfine transition complex of C$_2$H emission from the V4046 Sgr disk (we did not model the 262.064 GHz transition complex but, given its similar excitation conditions, we would expect similar results). The resulting best-fit radial column density profile and vertical and radial distributions of the C$_2$H emitting region are illustrated in Fig.~\ref{fig:C2Hmodel}. The parameters of the best-fit model are $Z_1 = 0.5$, $Z_2 = 1.5$ --- i.e., the emitting layer is confined to vertical disk column densities of between $\sim5\times10^{21}$ cm$^{-2}$ and $\sim5\times10^{22}$ cm$^{-2}$ --- with inner and outer radii $R_{in} = 30$ AU and $R_{out}=130$ AU. The uncertainties in the former (vertical) layer parameters ($Z_1, Z_2$) are of order 0.5 (i.e., a factor $\sim$3 uncertainty in vertical disk column density), and the uncertainties in the latter (radial) emission region boundaries are of order $\sim$5 AU. We find a model surface density power-law break point of $R_1 = 100$ AU, with a radial power-law slope of $p_1 = +0.6$ between 30 AU and 100 AU and a (much steeper) slope of $p_2 = +1.8$ between 100 AU and 130 AU. 
In the model, the C$_2$H column density ($N_{\rm C_2H}$) behaves essentially as a step function at $R_{in} = 30$ AU, increasing from $\sim$10$^6$ cm$^{-2}$ to $\sim$10$^{14}$ cm$^{-2}$ over the (unresolvable) span of just a few AU. $N_{\mathrm C_2H}$ then slowly increases to $\sim$$2\times10^{14}$ cm$^{-2}$ at $R_1 = 100$ AU before sharply increasing over the outer $\sim$30 AU in radius, to $N_{\rm C_2H} \sim 3.3\times10^{14}$ cm$^{-2}$ at the outer cutoff radius $R_{out}=130$ (Fig.~\ref{fig:C2Hmodel}, left). 

\begin{figure}[htbp]
\begin{center}
\includegraphics[width=3in]{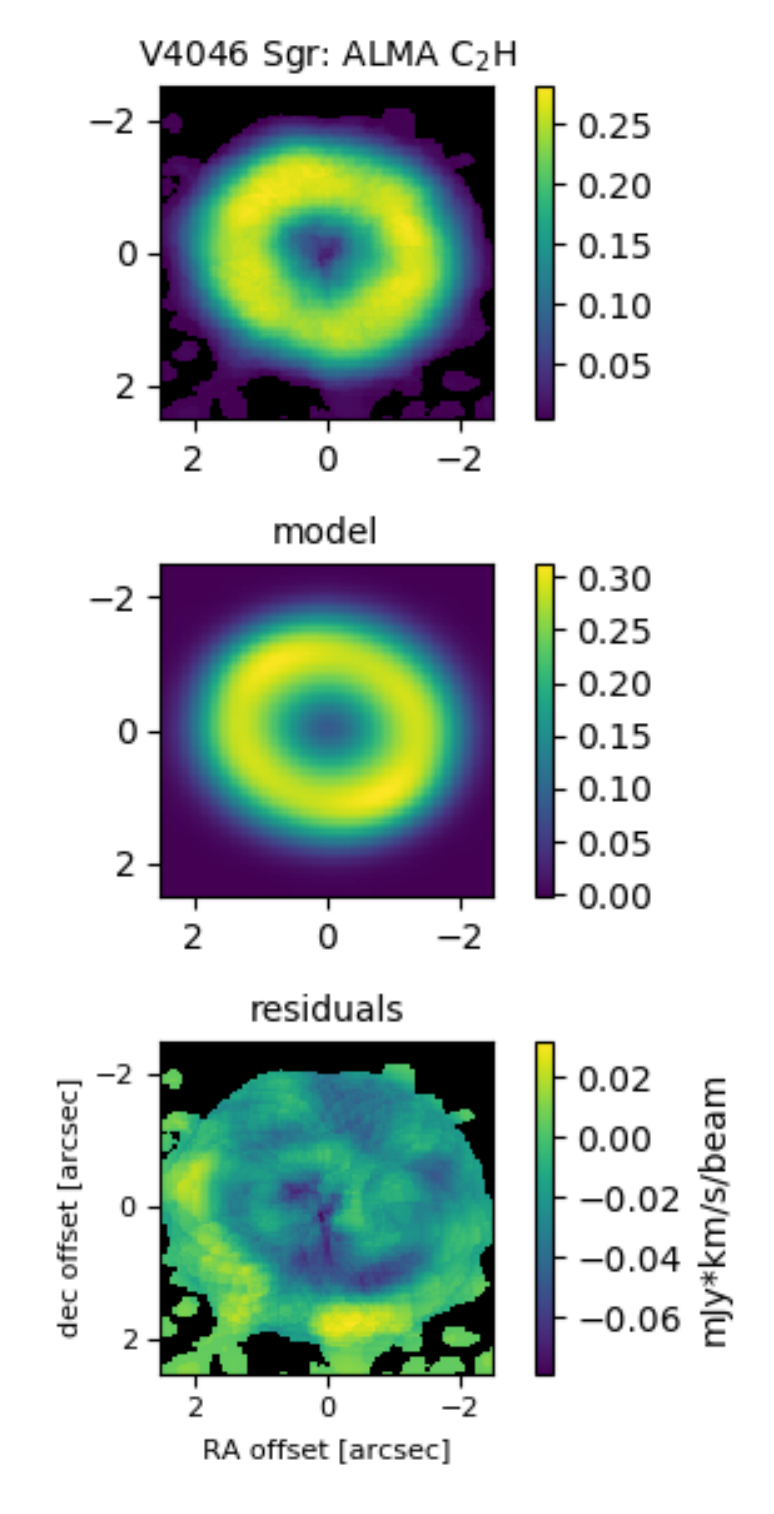}
\includegraphics[width=3in]{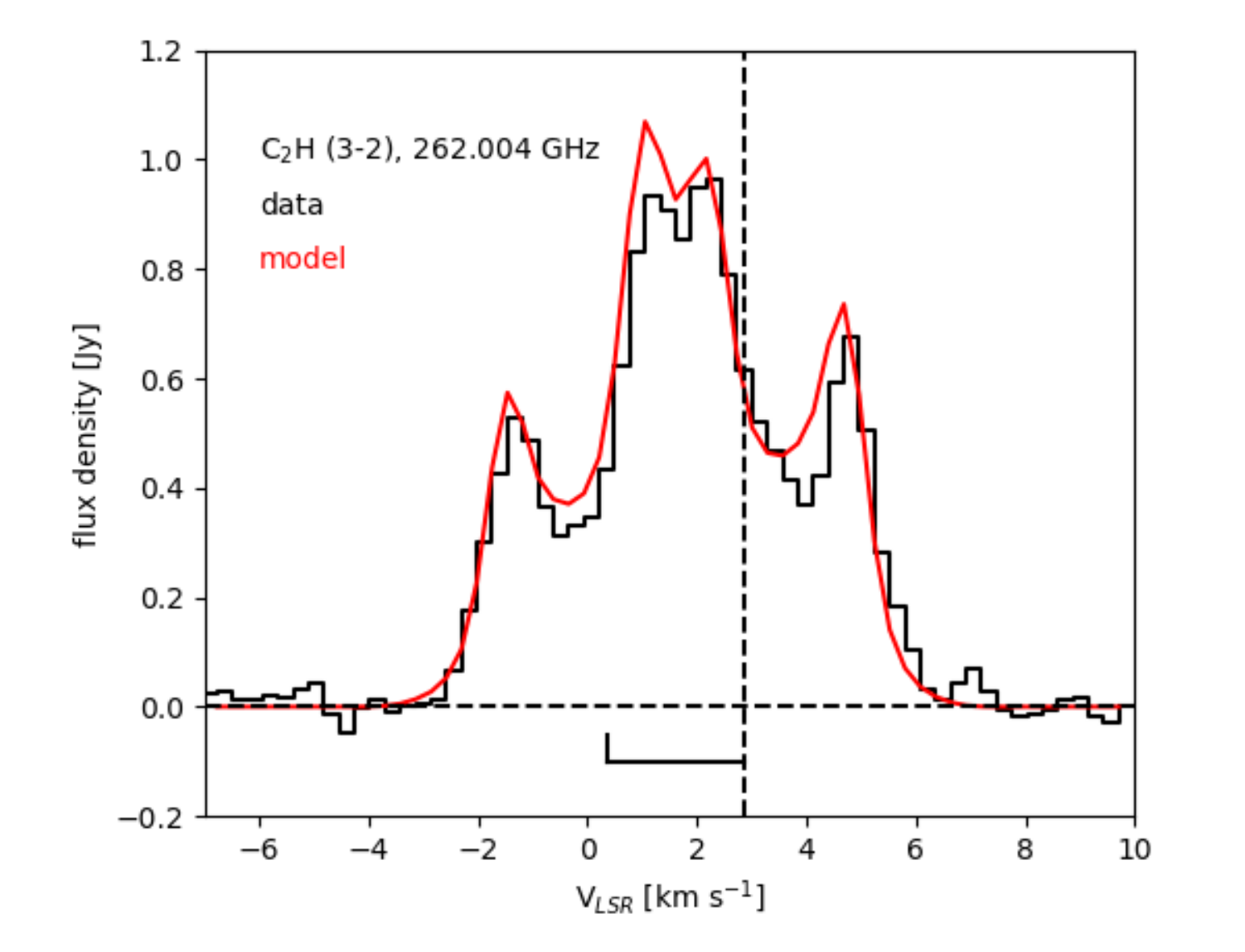}
\caption{Left: comparison of ALMA moment 0 image of C$_2$H 262.004 GHz emission from V4046 Sgr (top panel) with the moment 0 image obtained from the best-fit model (center panel); the residuals of the fit are shown in the bottom panel. Right: comparison of observed and model line profiles of C$_2$H 262.004 GHz emission, with the systemic velocity of V4046 Sgr (vertical dashed line) and the velocity offset corresponding to the hyperfine splitting of the C$_2$H transition indicated.
}
\label{fig:compareC2H}
\end{center}
\end{figure}

In Fig.~\ref{fig:compareC2H}, we compare the ALMA C$_2$H moment 0 image and line profile for V4046 Sgr with the corresponding image and line profile obtained from the best-fit model. In the latter Figure, the ALMA and model image data have been reconstructed with a \verb+robust+ value of 2.0, resulting in $\sim$0.8$''$ resolution images. It is evident that the modeling procedure has yielded a reasonable match to the ALMA C$_2$H image, as intended. However we caution that, as we have not employed a rigorous parameter space study, the robustness and uniqueness of this particular model fit, as well as the precise uncertainties in the various model parameters, are difficult to assess.

The ALMA C$_2$H modeling results illustrated in Fig.~\ref{fig:C2Hmodel} (left) supercede previous extrapolations of (much larger) $N_{\rm C_2H}$ in the V4046 Sgr disk, based on low S/N single-dish (APEX) spectroscopy \citep{Kastner2014}. Given the kinetic temperature regime in the emitting region ($T \sim 40$ K; Fig.~\ref{fig:C2Hmodel}, right), the model $N_{\rm C_2H}$ values indicate that the C$_2$H emission is optically thin; this inference is consistent with the fact that the measured ratio of total line intensities in the 262.004 GHz and 262.064 GHz hyperfine complexes, 1.37$\pm$0.1, is in good agreement with the theoretical ratio of 1.41 \citep{Ziurys1982}. 

In terms of disk scale height, the best-fit model C$_2$H emitting layer corresponds to the region between $z/r \sim 0.1$ and $z/r \sim 0.3$ over the radial range $\sim$30 AU to $\sim$130 AU (Fig.~\ref{fig:C2Hmodel}, right). This (relatively deep) vertical emitting region position contrasts with the surface layer position we inferred for the TW Hya disk, i.e.,  $z/r \sim 0.5$ (corresponding to far smaller vertical disk column densities, in the range $\sim$(5--8)$\times10^{18}$ cm$^{-2}$). However, as noted, we based our inference that the C$_2$H lies near the TW Hya disk surface on the apparently subthermal excitation of C$_2$H in that  disk \citep{Kastner2015}. In contrast to TW Hya, our empirical modeling of the integrated intensity image and line profile of C$_2$H emission from V4046 Sgr provides more direct constraints on the scale height of its emitting C$_2$H. Our results for V4046 Sgr hence suggest that the C$_2$H emitting layer may lie deeper within the TW Hya disk than we inferred previously.  However, we caution that the vertical density and temperature structures of the V4046 Sgr and TW Hya disks, and protoplanetary disks more generally, are subject to large uncertainties.

The vertical position of the emitting layer within V4046 Sgr is somewhat deeper in the disk than, but is not necessarily inconsistent with, the vertical C$_2$H layer position predicted by models in which production of C$_2$H is irradiation-driven; in particular, \citet{Walsh2010} find that the C$_2$H abundance should peak at $z/r \approx 0.3$. On the other hand, the low temperature ($\sim$20 K) and high densities ($>10^{7}$ cm$^{-3}$) of the disk layers corresponding to the region between $z/r \approx 0.1$ and $z/r \approx 0.3$ would imply that pure gas-phase production, perhaps enhanced by a large C/O ratio in the outer disk, is responsible for the large inferred C$_2$H column densities within the emission ring \citep[see discussions in][]{Kastner2015,Bergin2016}. Indeed, whereas pure gas-phase, deep-disk-layer production mechanisms appear to be able to generate ring-like C$_2$H emission morphologies \citep[][]{Henning2010}, surface layer irradiation production mechanisms --- such as photodesorption of hydrocarbon-coated grains or photodestruction of small grains --- may require {\it ad hoc} assumptions such as inner disk shadowing \citep[][]{Kastner2015}. Furthermore, under either (irradiation or pure gas-phase) production scenario, the ``cusp'' of high C$_2$H abundance at the outer edge of the C$_2$H ring that is required by our best-fit empirical model (in the form of an abrupt steepening of the power-law slope near $R_{out}$; Fig.~\ref{fig:C2Hmodel}, left) appears to be very difficult to explain. Observations of C$_2$H emission from other disks that are viewed at higher inclinations and are similarly nearby and well-resolved by ALMA \citep[e.g., T Cha;][]{Huelamo2015}, in combination with additional theoretical efforts aimed at better understanding C$_2$H production, are essential if we are to pinpoint the processes that lead to large abundances of C$_2$H within evolved protoplanetary disks.

\section{Summary}

We have presented a library of ALMA molecular line and continuum images of the circumbinary disk orbiting
V4046 Sgr, obtained during the course of three ALMA programs carried out in Cycles 2 and 3. All of these ALMA line surveys of V4046 Sgr have been undertaken in the 1.1--1.4 mm wavelength range (ALMA Bands 6 and 7) with ALMA antenna configurations involving maximum baselines of several hundred meters, yielding subarcsecond-resolution images in more than a dozen molecular species and isotopologues. Collectively, the resulting subarcsecond ALMA molecular line  images of V4046 Sgr serve to elucidate, on linear size scales of $\sim$30--40 AU, the chemical structure of an evolved, circumbinary, protoplanetary disk. 

The molecules CO and HCN and their isotopologues display centrally peaked velocity-integrated line intensity morphologies. The radial profiles of the intensity ratios of CO isotopologue emission serve to constrain the opacities in the $2 \rightarrow 1$ transitions of $^{12}$CO and $^{13}$CO. We find that both lines are optically thick throughout the disk. Their near-constant ratio of $\sim$2.5 across the disk then indicates that the $^{13}$CO emission arises from cooler, deeper disk layers than the $^{12}$CO emission, consistent with the predictions of irradiated disk models.

The integrated intensity of line emission from relatively complex nitriles (HC$_3$N, CH$_3$CN), deuterated molecules (DCN, DCO$^+$), hydrocarbons (as traced by C$_2$H), and potential CO ice line tracers (N$_2$H$^+$, and H$_2$CO) appears as a sequence of sharp and diffuse rings. The compact HC$_3$N and CH$_3$CN molecular emission regions appear as rings with dimensions similar to those of the central continuum emission and scattered-light rings within the V4046 Sgr disk. This correspondence suggests that the production of gas-phase HC$_3$N and CH$_3$CN may be driven, at  least in part, by photodesorption of organic ices from the surfaces of dust grains. The sequence of increasing radius of peak intensity represented by DCN, DCO$^+$, N$_2$H$^+$, and H$_2$CO is qualitatively consistent with the expected decline of midplane gas temperature with increasing disk radius, although the precise radial onset of midplane CO freezeout is difficult to ascertain given the lack of a clearly defined central hole in the N$_2$H$^+$ ring.

We have conducted empirical modeling of the C$_2$H emission ring so as to ascertain its radial and vertical extent within the V4046 Sgr disk.
We find that the sharp inner edge of C$_2$H emission lies at $\sim$30 AU, roughly coincident with the outer radius of the continuum (large-grain) and cyanide group emission, and the equally abrupt outer edge of C$_2$H lies at $\sim$130 AU. The modeling indicates that this C$_2$H ring lies deep within the disk, at scale heights of between $z/r \sim 0.1$ and $z/r \sim 0.3$. The relatively deep vertical position of this layer would appear to favor pure gas-phase C$_2$H production mechanisms over stellar high-energy radiation-driven production of C$_2$H, although the latter production mechanism is not ruled out. Either mechanism would be further enhanced by C enrichment (relative to O) in the outer disk. 

There is clearly far more to be learned about the V4046 Sgr disk via ALMA observations. Higher-resolution observations of its  mm-wave continuum and CO emission, complex nitrile emission, and HCN isotopologue emission have been carried out or are forthcoming (during Cycle 5).  However, there have been no successful observations of CN to date, and none are presently scheduled, leaving open key questions as to the production of CN and its utility as a tracer of cosmic nitrogen isotope ratios \citep{HilyBlant2017}.  More generally, the molecular ``anatomical'' study of the V4046 Sgr disk presented here should serve to motivate additional subarcsecond ALMA molecular line imaging surveys of similarly evolved and nearby protoplanetary disks spanning a range of inclinations, as well as further detailed chemical modeling. Such efforts should address the many uncertainties in protoplanetary disk physical and chemical structure and molecular production pathways touched on by the present study, such as the abundance distributions and (hence) production and destruction mechanisms of C$_2$H, nitriles, and other potentially robust tracers of disk irradiation and gas-grain processes. 

{\it This paper makes use of data from ALMA programs ADS/JAO.ALMA\#2013.1.00226.S (PI: K. \"{O}berg), \#2015.1.00671.S (PI: J. Kastner), and \#2015.1.00678.S (PI: C. Qi). ALMA is a partnership of ESO (representing its member states), NSF (USA) and NINS (Japan), together with NRC (Canada), NSC and ASIAA (Taiwan), and KASI (Republic of Korea), in cooperation with the Republic of Chile. The Joint ALMA Observatory is operated by ESO, AUI/NRAO and NAOJ. This research was supported by NASA Exoplanets program grant NNX16AB43G to RIT. Additional support for this research was provided to J.H.K., during his 2016-17 sabbatical, by the Universit\'{e} Grenoble Alpes and Institut de Plan\'{e}tologie et d'Astrophysique de Grenoble, by the Merle A. Tuve Senior Fellowship at the Carnegie Institution's Department of Terrestrial Magnetism, and by a Smithsonian Institution Visitor's Fellowship with the Radio and Geoastronomy Division of the Harvard-Smithsonian Center for Astrophysics. The authors thank Dary Ruiz-Rodriguez for useful suggestions.}

%\bibliography{refs}

\begin{table}
%\centering
\caption{\sc Subarcsecond ALMA Molecular Line Imaging of V4046 Sgr}
\label{tbl:LibraryImages}

\tiny

\vspace{.2in}

\begin{tabular}{ccccccccc}
\hline
\hline
Program & Species & Trans. & $\nu_0$ & $E_L/k$ & beam (PA)$^a$  & $\delta v$$^b$ & rms$^b$ & $I$$^c$\\
              & & & (GHz) & (K) & & (km s$^{-1}$) & (mJy beam$^{-1}$) & (Jy km s$^{-1}$) \\
\hline
2013.1.00226.S & DCO$^+$ & $J=3\rightarrow 2$ & 216.11258 & 10.37 & $0.76''$$\times$$0.49''$ ($-$87.6$^\circ$) & 0.17 & 0.5& 0.51$\pm$0.02 \\
2013.1.00226.S & C$^{18}$O & $J=2\rightarrow 1$ & 219.56036 & 5.29 & $0.81''$$\times$$0.48''$ ($-$85.5$^\circ$) & 0.17 & 3.7& 2.11$\pm$0.09\\
2013.1.00226.S & $^{13}$CO & $J=2\rightarrow 1$ & 220.39868 & 5.27 & $0.81''$$\times$$0.48''$ ($-$86.2$^\circ$) & 0.17 & 6.3& 8.44$\pm$0.45\\
%2015.1.00671.S & CN & $N=2\rightarrow 1$$^d$ & 226.65956 & 5.45 & {\bf [???]} & 0.65 & \\
%2015.1.00671.S & CN & $N=2\rightarrow 1$$^e$ & 226.87478 & 5.45 & {\bf [???]} & 0.65 & \\
%2015.1.00226.S & $^{12}$CO & $J=2\rightarrow 1$ & 230.538 & 5.53 & $0.72''$$\times$$0.44''$ ($+$84.6$^\circ$)& 0.17 & 5.6& 28.65 \\
2015.1.00226.S & $^{12}$CO & $J=2\rightarrow 1$ & 230.538 & 5.53 & $0.79''$$\times$$0.48''$ ($-$90.0$^\circ$)& 0.17 & 5.6& 29.30$\pm$0.54\\
2013.1.00226.S & HC$_3$N & $J=27\rightarrow 26$ & 245.60632 & 153.2 & $0.55''$$\times$$0.46''$ ($+$77.9$^\circ$) & 0.15 & 1.8& 0.33$\pm$0.04\\
2013.1.00226.S & CH$_3$CN & $J=14\rightarrow 13$ & 257.50756 & 108.9 & $0.53''$$\times$$0.43''$ ($+$77.0$^\circ$) & 0.14 & 1.8& 0.43$\pm$0.04\\
2013.1.00226.S & HC$^{15}$N & $J=3\rightarrow 2$ & 258.1570 & 12.39 & $0.53''$$\times$$0.43''$ ($+$75.9$^\circ$) & 0.14 & 1.6& 0.45$\pm$0.04\\
2013.1.00226.S & H$^{13}$CN & $J=3\rightarrow 2$ & 259.01180 & 12.43 & $0.52''$$\times$$0.43''$ ($+$78.1$^\circ$) & 0.14 & 2.1& 0.74$\pm$0.05\\
2013.1.00226.S & H$^{13}$CO$^+$ & $J=3\rightarrow 2$ & 260.25534 & 12.49 & $0.52''$$\times$$0.43''$ ($+$77.5$^\circ$) & 0.14 & 2.0& 0.25$\pm$0.14\\
2015.1.00671.S & C$_2$H & $N=3\rightarrow 2$ & 262.00423$^d$ & 12.58 & $0.56''$$\times$$0.42''$ ($-$76.4$^\circ$)& 0.56  & 6.9& 4.37$\pm$0.24\\ 
2015.1.00671.S & C$_2$H & $N=3\rightarrow 2$ & 262.06484$^e$ & 12.58 & $0.61''$$\times$$0.41''$ ($-$76.1$^\circ$) & 0.56 & 7.3& 3.18$\pm$0.24\\
2015.1.00671.S & HCN & $J=3\rightarrow 2$ & 265.88618 & 12.76 &  $0.53''$$\times$$0.41''$ ($-$74.0$^\circ$) & 0.28 & 7.9 & 8.61$\pm$0.40\\
2015.1.00678.S & N$_2$H$^+$ & $J=3\rightarrow 2$ & 279.5117 & 13.41 & $0.59''$$\times$$0.49''$ ($-$82.7$^\circ$) & 0.27 & 5.8 & 4.16$\pm$0.52\\
2015.1.00678.S & DCN & $J=4\rightarrow 3$ & 289.64524 & 14.49 & $0.55''$$\times$$0.45''$ ($-$77.5$^\circ$) & 0.50 & 4.6 & 0.68$\pm$0.21\\
2015.1.00678.S & H$_2$CO & 4(0,4)$\rightarrow$3(0,3) & 290.62341 & 14.57 & $0.55''$$\times$$0.44''$ ($-$79.7$^\circ$) & 0.50 & 5.1 & 1.98$\pm$0.41\\
\hline
\end{tabular}

NOTES:\\
a) Major and minor axes and position angle of synthesized beam. Briggs
\verb+robust+ parameter set to 0.0 for C$_2$H and HCN; 0.5 for DCO$^+$, $^{13}$CO
C$^{18}$O, HC$_3$N, CH$_3$CN, HC$^{15}$N, H$^{13}$CN, and H$^{13}$CO$^+$; 1.0 for
$^{12}$CO; and 2.0 for DCN, H$_2$CO, and N$_2$H$^+$. \\
b) Channel width and rms image noise per channel. \\
c) Integrated flux in velocity-integrated (moment 0)
image. \\
%d) Centered on $J =3/2 \rightarrow 1/2, F =5/2 \rightarrow 3/2$ transition, but includes flux from neighboring HF transitions. \\
%e) Centered on $J =5/2 \rightarrow 3/2, F =7/2 \rightarrow 5/2$ transition, but includes flux from neighboring HF transitions. \\
d) Frequency of $J =7/2 \rightarrow 5/2, F =4 \rightarrow 3$
transition; includes flux from $F =4 \rightarrow 3$ transition at
262.00640 GHz. \\
e) Frequency of $J =5/2 \rightarrow 3/2, F =3 \rightarrow 2$
transition; includes flux from $F =2 \rightarrow 1$ transition at
262.06733 GHz. \\

\end{table}

\begin{table}
%\centering
\caption{\sc ALMA mm-wave Continuum Measurements of V4046 Sgr}
\label{tbl:Continuum}

\footnotesize

\vspace{.2in}

\begin{tabular}{cccc}
\hline
\hline
Program & $\nu$ & beam (PA)$^a$  & flux$^b$ \\
              & (GHz) & & (mJy) \\
\hline
%2015.1.00671.S & 228.0 & {\bf TBR} & {\bf TBR} \\
2013.1.00226.S & 235 & $0.70''$$\times$$0.43''$ ($+$84$^\circ$) & 338$\pm$34 \\
2013.1.00226.S & 251 & $0.52''$$\times$$0.43''$ ($+$82$^\circ$) & 353$\pm$35 \\
2015.1.00671.S & 264 & $0.69''$$\times$$0.49''$ ($-$77$^\circ$)  &  327$\pm$33 \\
2015.1.00678.S & 284 & $0.38''$$\times$$0.29''$ ($-$73$^\circ$)  &  472$\pm$47 \\
\hline
\end{tabular}

NOTES:\\
a) Major and minor axes and position angle of synthesized beam.\\
b) Integrated fluxes above 3 $\sigma$ level in continuum images;
listed errors assume
(10\%) systematic calibration uncertainties dominate over
pixel-to-pixel rms uncertainties.

\end{table}

\end{document}